\documentclass[twocolumn]{aastex631}

\usepackage{multirow}

\usepackage[utf8]{inputenc}
\usepackage{natbib}
\usepackage{graphicx}
\usepackage{longtable}
\usepackage{enumerate}
\usepackage{amsmath}
\usepackage{hyperref}
\usepackage{CJK}
\usepackage[title]{appendix}
\usepackage{rotating}
\usepackage{longtable}

\usepackage{amsmath}

\def\msun{\ifmmode {\rm M}_{\mathord\odot}\else $M_{\mathord\odot}$\fi}
\def\rsun{\ifmmode {\rm R}_{\mathord\odot}\else $R_{\mathord\odot}$\fi}
\def\lsun{\ifmmode {\rm L}_{\mathord\odot}\else $L_{\mathord\odot}$\fi}

\def\co{$^{12}$CO}
\def\c18o{C$^{18}$O}
\def\h2{H$_{2}$}
\def\13co{$^{13}$CO}
\def\n2hp{$_{2}$H$^{+}$}

\def\radmc{{\sc radmc-3d}}
\def\cm2{cm$^{-2}$}
\def\cmc{cm$^{-3}$}

\newcommand{\kms}{km~s$^{-1}$}

\newcommand{\CASItD}{{\sc casi-3d}}

\def\orion{{\sc orion2}}
\def\deg{$^{\circ}$}

\shorttitle{}
\shortauthors{}

\begin{document}
\begin{CJK*}{UTF8}{gbsn}

\title{Application of Convolutional Neural Networks to Predict Magnetic Fields Directions in Turbulent Clouds}

\author[0000-0001-6216-8931]{Duo Xu}
\affiliation{Department of Astronomy, University of Virginia, Charlottesville, VA 22904-4235, USA}

\author[0000-0003-1964-970X]{Chi-Yan Law}
\affiliation{Department of Space, Earth \& Environment, Chalmers University of Technology, SE-412 96 Gothenburg, Sweden}
\affiliation{European Southern Observatory, Karl-Schwarzschild-Strasse 2, D-85748 Garching, Germany}

\author[0000-0002-3389-9142]{Jonathan C. Tan}
\affiliation{Department of Astronomy, University of Virginia, Charlottesville, VA 22904-4235, USA}
\affiliation{Department of Space, Earth \& Environment, Chalmers University of Technology, SE-412 96 Gothenburg, Sweden}

\email{xuduo117@virginia.edu}

\begin{abstract}
We adopt the deep learning method \CASItD\ (Convolutional Approach to Structure Identification-3D) to infer the orientation of magnetic fields in sub-/trans- Alfv\'enic turbulent clouds from molecular line emission. We carry out magnetohydrodynamic simulations with different magnetic field strengths and use these to generate synthetic observations. We apply the 3D radiation transfer code \radmc\ to model \co\ and \13co\ (J = 1-0) line emission from the simulated clouds and then train a \CASItD\ model on these line emission data cubes to predict magnetic field morphology at the pixel level. The trained \CASItD\ model is able to infer magnetic field directions with low error ($\lesssim 10^{\circ}$ for sub-Alfv\'enic samples and $\lesssim 30^{\circ}$ for trans-Alfv\'enic samples). We furthermore test the performance of \CASItD\ on a real sub-/trans- Alfv\'enic region in Taurus. The \CASItD\ prediction is consistent with the magnetic field direction inferred from {\it Planck} dust polarization measurements. We use our developed methods to produce a new magnetic field map of Taurus that has a three times higher angular resolution than the {\it Planck} map. 

\end{abstract}


\section{Introduction}
\label{Introduction}

Magnetic ($B$-) fields are ubiquitous in the Universe \citep[e.g.,][]{1999ApJ...520..706C,2017ARA&A..55..111H}. They are one of the major components regulating the motion and evolution of the interstellar medium (ISM) \citep{2012ARA&A..50...29C,2015MNRAS.450.4035F}. Although gravity and turbulence also play important roles in the formation of structures in the ISM \citep{1999ApJ...526..279P,2004ARA&A..42..211E,2007ARA&A..45..565M}, recent studies indicate that magnetic fields are also very important over a wide range of regimes and scales \citep{2009ApJ...704..891L,2014ApJ...789...82C,2019ApJ...871...98Z}. However, measuring $B$-fields is challenging. The observations of magnetic fields are divided into two main types of measurement. One is the plane-of-sky (POS) component, which is usually traced by polarized thermal dust emission \citep{1998ApJ...502L..75R,2016A&A...586A.138P}, starlight polarization \citep{1951ApJ...114..206D,2002ApJ...564..762F}, and synchrotron emission \citep{1982A&A...105..192B,2012ApJ...761L..11J}. The other is the line-of-sight (LOS) component, which is normally resolved by Zeeman splitting \citep{1986ApJ...301..339T,2010ApJ...725..466C} and Faraday rotation \citep{1966MNRAS.133...67B,2022A&A...657A..43H}. 

Although we have multiple approaches to trace POS $B$-fields, their study remains challenging. For example, starlight polarization is limited by the lines of sight along which stars
are present and detectable. The polarized thermal dust emission is considered to be caused by radiative torques (RATs) produced by anisotropic radiation flux with respect to the magnetic field \citep{2008MNRAS.388..117H}, which imposes other conditions related to the presence of such flux.
%requires a relatively high column density ($A_V\sim 10$). 
Moreover, both methods only trace the projected POS magnetic field direction and are not capable of distinguish the POS $B$-fields for different gas components on the LOS.

Approaches to studying $B$-fields based on theories of interstellar magnetohydrodynamic (MHD) turbulence and turbulent
reconnection have also been developed \citep{1995ApJ...438..763G,1999ApJ...517..700L}. \citet{1995ApJ...438..763G} examined the effects of the interaction among shear Alfv\'en waves and found that turbulent eddies become elongated along the magnetic field direction in the limit of strong Alfv\'en turbulence, where the perturbed velocity is much smaller than the Alfv\'en velocity, i.e., sub-Alfv\'enic. 
%jct - what does strong Alfven turbulence mean? i.e., what values of Alfven mach number?
%jct - I think we need some text to better explain what VGT is - please check this:
Based on this crucial principle, a method to trace the direction of magnetic fields by using spectroscopic data has been developed, i.e., the velocity channel gradient technique \citep[VGT \footnote{  {There are different acronyms or naming of the same technique in different literature studies, including  Velocity channel Gradient(VchG) in \citet{2018ApJ...853...96L} and `gradients of surface brightness within thin velocity slices' in \citet{2020MNRAS.496.4546H}. Also, the original velocity gradient technique \citep{2017ApJ...835...41G} was applied on the velocity centroid map instead of thin-slice velocity channel maps. Here we will use the acronym VGT throughout the work to preserve the original spirit of using velocity gradients as a means of magnetic field tracing.}},][]{2018ApJ...853...96L,2020MNRAS.496.4546H}.  
 { VGT examines maps of line emission from the gas in small velocity ranges and measures the gradient in intensity of this emission in small patches, i.e., ``sub-blocks''. Then the plane of the sky magnetic field direction is assumed to be in the direction perpendicular to this gradient.}
 %There are other short forms of the same techniques, including Velocity channel Gradient (VchG) or 'intensity gradient of thin-slice velocity channels' \citep{2020MNRAS.496.4546H}}

VGT has been examined in numerical simulations and synthetic spectroscopic data \citep{2017ApJ...837L..24Y,2018ApJ...853...96L} and tested on observational data \citep{2019NatAs...3..776H,2022MNRAS.510.4952L}. 
%jct - this does not make sense to me... why 50x50? And what is a "pixel"? It should simply require 2 or 3 beam scales
However, \citet{2017ApJ...837L..24Y} pointed out that the gradient calculation requires sub-block averaging 
%(about minimum of $50\times 50$ pixels) 
to obtain a robust result
%jct - I do not understand the point of the next sentence
%In addition, as the sub-block size increases, the mean gradient direction becomes more well-defined.
%Unfortunately, 
and this averaging process limits the resolution that can be achieved to determine $B$-field morphology.
Furthermore, it remains somewhat unclear about the accuracy of the VGT method in different situations of application. { For example, \citet{2019ApJ...874..171C} pointed out that the gas emission in the velocity channel maps, such as the H\,{\sc i} 21 cm line, is not dominated by velocity fluctuations, but rather by density fluctuations, which implies that the fundamental assumption of VGT is not valid any more. In addition, when self-gravity is important, then the velocity structures will be influenced by additional effects that would lead to VGT analysis yielding inaccurate results \citep[e.g.,][]{2022ApJ...928..132L}. Neverthelesse, while the importance of velocity fluctuations on the gas emission in thin velocity channels is still under debate, a number of simulations have demonstrated that magnetic fields can play an important role in shaping the morphology of gas emission in velocity channel maps \citep{2013ApJ...774..128S,2016ApJ...833...10I,2017A&A...607A...2S}. Moreover, an increasing amount of observational evidence shows that the morphology of the ISM is regulated by magnetic fields, including the atomic cold neutral medium \citep{2014ApJ...789...82C,2015PhRvL.115x1302C} and molecular gas \citep{2019A&A...629A..96S,2020MNRAS.496.4546H}, including via polarized dust emission \citep{2017A&A...603A..64S}.}
%reduces the resolution by a factor of a few. 
Consequently, we introduce a deep learning method, i.e., via convolutional neural networks (CNNs), to examine the morphology of spectroscopic data and infer the magnetic field directions in simulations and observations of the ISM. 

%jct - ok

CNNs have gained increasing popularity among astronomers and are widely used for a variety of tasks, including galaxy morphology prediction \citep{2015MNRAS.450.1441D}, exoplanet detection \citep{2018MNRAS.474..478P} and stellar feedback bubble identification \citep{ 2019ApJ...880...83V}. CNNs learn the morphology and extract the feature of objects by applying a series of kernels to convolve with the input data. CNNs are proficient at identifying and outlining objects from the input data. \citet{2020ApJ...890...64X,2020ApJ...905..172X} developed a Convolutional Approach to Structure Identification-3D (\CASItD) based on CASI-2D \citep{2019ApJ...880...83V} to identify stellar feedback bubbles and protostellar outflows in position-position-velocity (PPV) molecular line spectral cubes.  \citet{2020ApJ...890...64X,2020ApJ...905..172X} trained CASI-3D on synthetic molecular line spectral cubes and applied it to real observations. They identified all previously visually identified feedback structures in nearby molecular clouds \citep{2022ApJ...926...19X}. \citet{2021A&A...652A.143B} applied CNNs to infer the radial component and the transverse magnetic field of the Sun from 2D photospheric continuum images and achieved a low uncertainty $\sim10\%$. This demonstrates the ability of CNNs to infer magnetic fields based on the morphology of emission and/or absorption. \citet{2019ApJ...882L..12P} applied CNNs to distinguish between sub-Alfv\'enic and super-Alfv\'enic turbulence from the Fourier phase space of a density slice from simulations and achieved a high accuracy $>98$\%. This work indicates that there is a significant amount of information in images, e.g., in Fourier phase space, compared to just a single power spectrum. The results from \citet{2019ApJ...882L..12P} and \citet{2021A&A...652A.143B} indicate that CNNs are able to retrieve more information, e.g., magnetic fields information at the pixel level, from PPV cubes.

In this paper, we adopt the deep learning method \CASItD\ to infer the orientation of magnetic fields in sub-/trans- Alfv\'enic turbulent clouds from molecular line emission. We describe \CASItD\ and how we generate the training set from synthetic observations in Section~\ref{Data and Method}. Here we also introduce the CO observations and {\it Planck} dust polarization data. In Section~\ref{Results}, we evaluate our CNN models in predicting the orientation of magnetic fields on synthetic observations and present the performance of CNN models on real observations. We summarize our results and conclusions in Section~\ref{Conclusions}.

\section{Data and Method}
\label{Data and Method}

\subsection{Magnetohydrodynamics Simulations}
\label{Magnetohydrodynamics Simulations}
We conduct ideal MHD simulations with \orion\ \citep{2012ApJ...745..139L} to model sub- and trans-Alfv\'enic turbulent clouds. The simulation box is $5\times5\times5$~pc$^3$ with periodic boundary
conditions and without self-gravity. The magnetic field is initialized along the $z$ direction. We treat the gas as an isothermal ideal gas with an initial cloud temperature of 10 K. The three-dimensional Mach number is 10.5, which places the simulated cloud on the line width-size relation, $\sigma _{\rm 1D}=0.72 R^{0.5}_{\rm pc}$ \kms \citep{2007ARA&A..45..565M}. The calculations use a base grid of 256$^3$. We conduct simulations with two sets of virial parameters $\alpha_{\rm vir}=5\sigma_v^2 R/(GM)$: $\alpha_{\rm vir}=1$, which corresponds to a mean H-nuclei density
%jct - check this is correct for n_H, including if we assume n_He=0.1 n_H
of $n_{\rm H}=1046$~\cmc\ and total mass of 3767~\msun; and $\alpha_{\rm vir}=2$, which corresponds to a mean H-nuclei density of $n_{\rm H}=523$~\cmc\ and total mass of 1884~\msun. Furthermore, for each value of $\alpha_{\rm vir}$
%jct - is this for each virial parameter?
we run two sets of simulations with different mass-to-flux ratios $\mu_{\Phi}=M_{\rm gas}/M_{\rm \Phi}=2\pi G^{1/2}M_{\rm gas}/(BL^2)$, $\mu_{\Phi}$= 1 and 2, which yields an Alfv\'en mach number between 0.6 to 1.7, indicating a sub-/trans-Alfv\'enic turbulent cloud. We initialize the density and velocity fields by driving the simulation gas for two Mach crossing times without gravity, but adding random large-scale perturbations with Fourier modes $1\le k \le 2$ \citep{1999ApJ...524..169M}. We list the physical properties of the simulations in Table~\ref{tab-simulation-turb}.

\begin{table}[]
  
\begin{center}
  \caption{Fitting Results of the Spatial Power Spectrum$^a$ \label{tab-simulation-turb}}
 \begin{tabular}{ccccc}
\hline
\hline
Model & $\alpha_{\rm vir}$   & $\mathcal{M}_{A}$ & $\mu_{\rm Phi}$  & $N_{\rm seed}$ \\
\cline{1-5} 
Turb1 & 1 & 1.24 &  2 & 2 \\ 
Turb2 & 1 & 0.62 &  1 & 2 \\
Turb3 & 2 &  1.75  &  2 & 2 \\
Turb4 & 2 & 0.87 &  1 & 2 \\
\cline{1-5} 
\multicolumn{5}{p{0.37\linewidth}}{Notes:}\\
\multicolumn{5}{p{0.37\linewidth}}{$^a$ Model name, virial parameters, Alfv\'en mach number, mass-to-flux ratio, and the number of different turbulent driving patterns.}
\end{tabular}%
\end{center}
\end{table}%

\subsection{Training Sets}
\label{Training Sets}

\begin{figure*}[hbt!]
\centering
\includegraphics[width=0.68\linewidth]{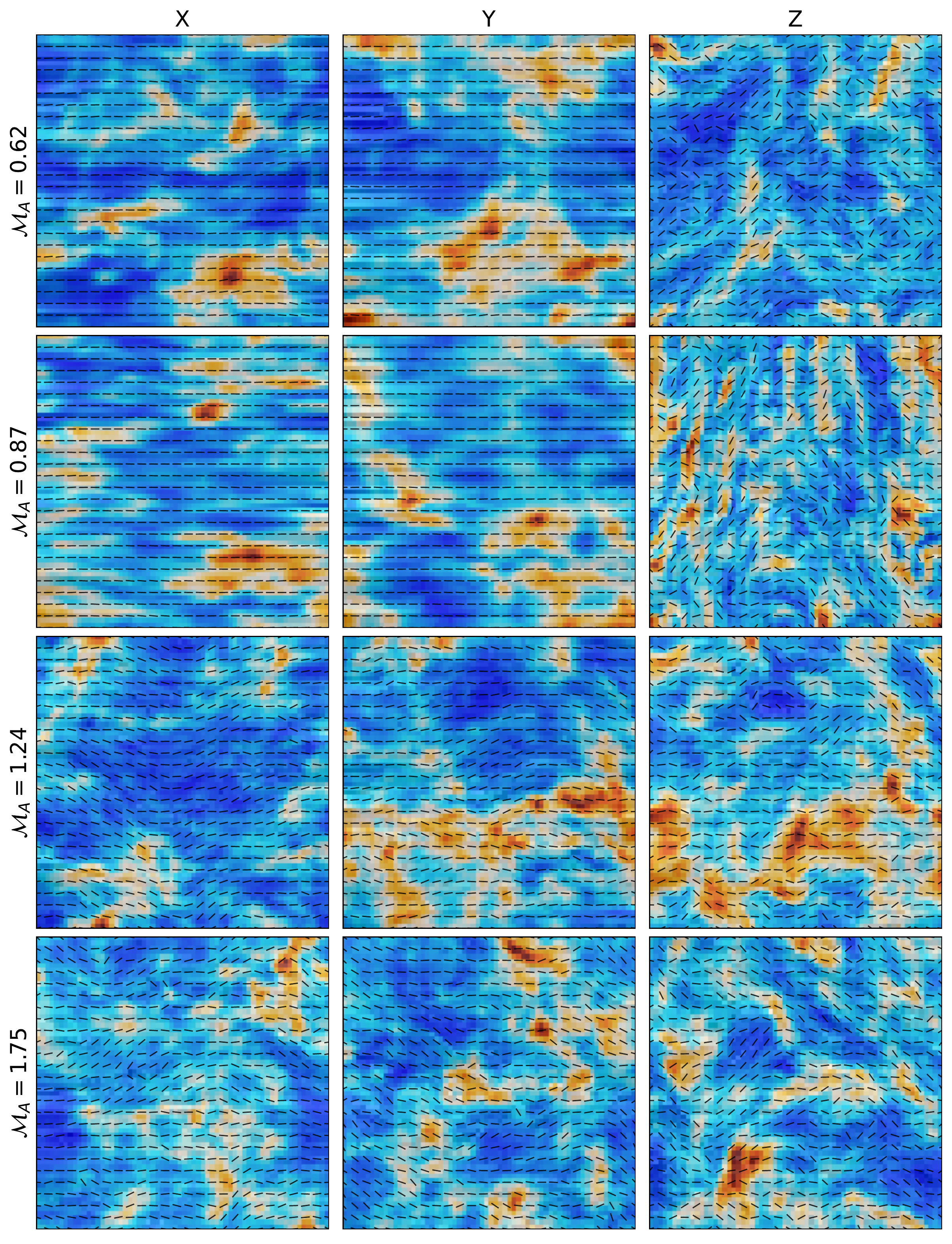}
\caption{Synthetic \13co (1-0) integrated intensity overlaid with the direction of magnetic fields. Three columns are three different viewing directions. Four rows represent simulations with different Alfv\'en mach numbers, as labelled.}
\label{fig.turb-synthetic-co13-bfield-raw}
\end{figure*}

\begin{figure*}[hbt!]
\centering
\includegraphics[width=0.68\linewidth]{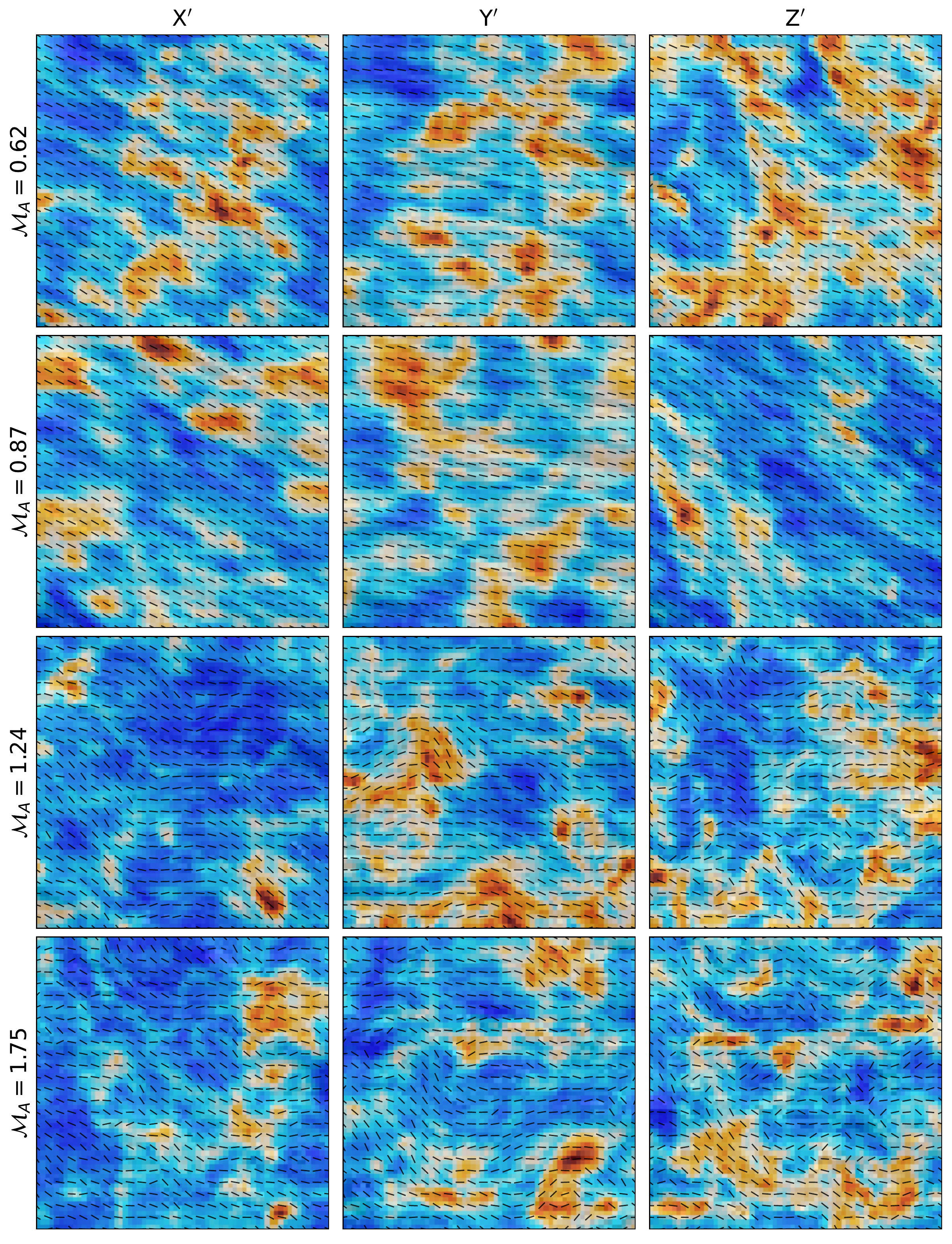}
\caption{Same as Figure~\ref{fig.turb-synthetic-co13-bfield-raw}, but on a rotated simulation with Euler angles of (45\deg, 45\deg, 45\deg). }
\label{fig.turb-synthetic-co13-bfield-rot}
\end{figure*}

We apply the publicly available radiative transfer code \radmc\ \citep{2012ascl.soft02015D} to model \co\ ($J=1-0$) and \13co\ ($J=1-0$) line emission of the turbulent clouds. We adopt the simulation density, temperature and velocity distributions as the \radmc\ inputs. In the radiative transfer, we assume that \h2\ is the only collisional partner of CO. We take 62 as the fiducial abundance ratio between \co\ and \13co\ and $10^{-4}$ as
the abundance ratio between \co\ and \h2.

Figure~\ref{fig.turb-synthetic-co13-bfield-raw} shows the synthetic \13co\ (1-0) integrated intensity and the POS magnetic fields direction of the MHD simulation with different Alfv\'en mach numbers. Since the magnetic fields are initialized along the $z$ axis in these MHD simulations, the POS magnetic field directions viewed along the $z$ axis are less ordered. To reduce the impact of turbulence dominant pattern rather than that regulated by magnetic fields, we rotate the turbulent box following Euler angles ($\alpha$, $\beta$, $\gamma$) and conduct the radiative transfer from the new axes. We choose three sets of Euler angles (15\deg, 15\deg, 15\deg), (30\deg, 30\deg, 30\deg) and (45\deg, 45\deg, 45\deg). Figure~\ref{fig.turb-synthetic-co13-bfield-rot} shows the synthetic \13co\ integrated intensity and the POS magnetic fields directions of the MHD simulations rotated by Euler angles of (45\deg, 45\deg, 45\deg).

To enhance the diversity of physical and chemical conditions for the training set, we generate synthetic observations where both \co\ and \13co\ abundances are
reduced by a factor of 10. We also increase the training set by considering thin clouds with thicknesses between 0.7 and 5 pc using the same method in \citep{2020ApJ...890...64X}. We also conduct synthetic
observations with different physical scales, i.e., a “zoomed-in” synthetic observation with an image size of 2.5 pc $\times$ 2.5 pc, and 1.25 pc $\times$ 1.25 pc, and a “zoomed-out” synthetic observation with an image size of 5 pc $\times$ 5 pc. In addition to the different image sizes, we resample the synthetic observations with two different velocity resolutions: at low resolution with an interval of 0.25 \kms; and at high-resolution with an
interval of 0.125 \kms. Additionally, we rotate the images randomly from 0\deg\ to 360\deg\ and randomly shift the central velocity of the cubes from -3 to +3 \kms. Considering the optically thick emission of \co, we adopt \13co\ emission as the training set for \CASItD, which better outlines the morphology of the clouds. It is worth noting that the \CASItD\ model cares about the relative intensity of the data cube and does not require the data cube from a specific molecule. It is able to handle spectroscopic data from different molecules or atoms.

We then derive the magnetic field directions of the gas at each velocity channel from the simulation data. We first bin the LOS gas into velocity channels and calculate the mass-weighted magnetic field direction in each channel. The input data for \CASItD\ model is the PPV \13co\ data cube, and the target is the corresponding PPV magnetic field directions. In total we generate 11540 synthetic cubes: 6924 as a training set, 2308 as a test set and 2308 as a validation set.

\subsection{\CASItD: Inferring Orientation of Magnetic Fields}
\label{sec-CASItD}
In this section, we introduce a new \CASItD\ model to predict the orientation of magnetic fields from molecular line emission. We adopt the same CNN architecture, \CASItD, from \citep{2020ApJ...890...64X}. \CASItD\ is an autoencoder with both residual networks \citep{he2016deep} and a “U-net” \citep{ronneberger2015u}. We adopt the same hyperparameters as \citet{2020ApJ...890...64X,2020ApJ...905..172X}.

\subsection{Observations}
\label{Observations}

\subsubsection{\co\ and \13co\ Data}
\label{co and 13co data}

The \co\ ($J=1-0$) and \13co\ ($J=1-0$) lines were observed simultaneously in surveys of Taurus between 2002 and 2005 using the 13.7m Five College Radio Astronomy Observatory (FCRAO) Telescope \citep{2008ApJS..177..341N}. The \co\ and \13co\ maps are centered at $\alpha(2000.0)=04^{\rm h}32^{\rm m}44.6^{\rm s}$, $\delta(2000.0)=24^{\circ}25^{\prime}13.08^{\prime\prime}$ covering an area of 98 deg$^2$. The main beam of the antenna pattern has a FWHM of 45\arcsec\ for \co\ and 47\arcsec\ for \13co. The data are obtained on the fly (OTF), but they are resampled onto a uniform 23\arcsec\ grid \citep{2006AJ....131.2921R}. The Taurus data has a RMS antenna temperature of 0.28 K for \co\ and 0.125 K for \13co. There are 80 and 76 channels with 0.26 and 0.27~\kms\, spacing for \co\ and \13co, respectively. The velocity range of the Taurus data spans -5.1 to 14.9 \kms.

\subsubsection{{\it Planck} 353 GH Dust Polarization Map}
\label{Planck 353 GH Dust Polarization Map}
We adopt the data from the {\it Planck} 3rd Public Data Release \citep{2020A&A...641A..12P}. We infer the magnetic field orientation from the dust polarization angle:
%jct - what is arctan2?
\begin{align}
\label{phi-B}
\phi_{B} =\frac{1}{2} {\rm arctan2} (-U,Q)+\frac{\pi}{2},
\end{align}
where Q and U are the Stokes parameters of polarized dust emission, and ${\rm arctan2}$ is the two arguments arctangent function which returns the angle in the range (-$\pi$,$\pi$). The maps of Q and U are initially at 4$^{\prime}$.8 resolution in HEALPix
format with an effective pixel size of 1$^{\prime}$.07.

\section{Results}
\label{Results}

\subsection{Evaluation of \CASItD\ Performance on Synthetic Observations}
\label{Evaluate CAStD Performance on Synthetic Observation}
In this section, we use the synthetic data to assess how accurately magnetic field orientations can be determined from molecular line emission.

%jct - I would change the red color... perhaps to white or gray.
\begin{figure*}[hbt!]
\centering
\includegraphics[width=0.68\linewidth]{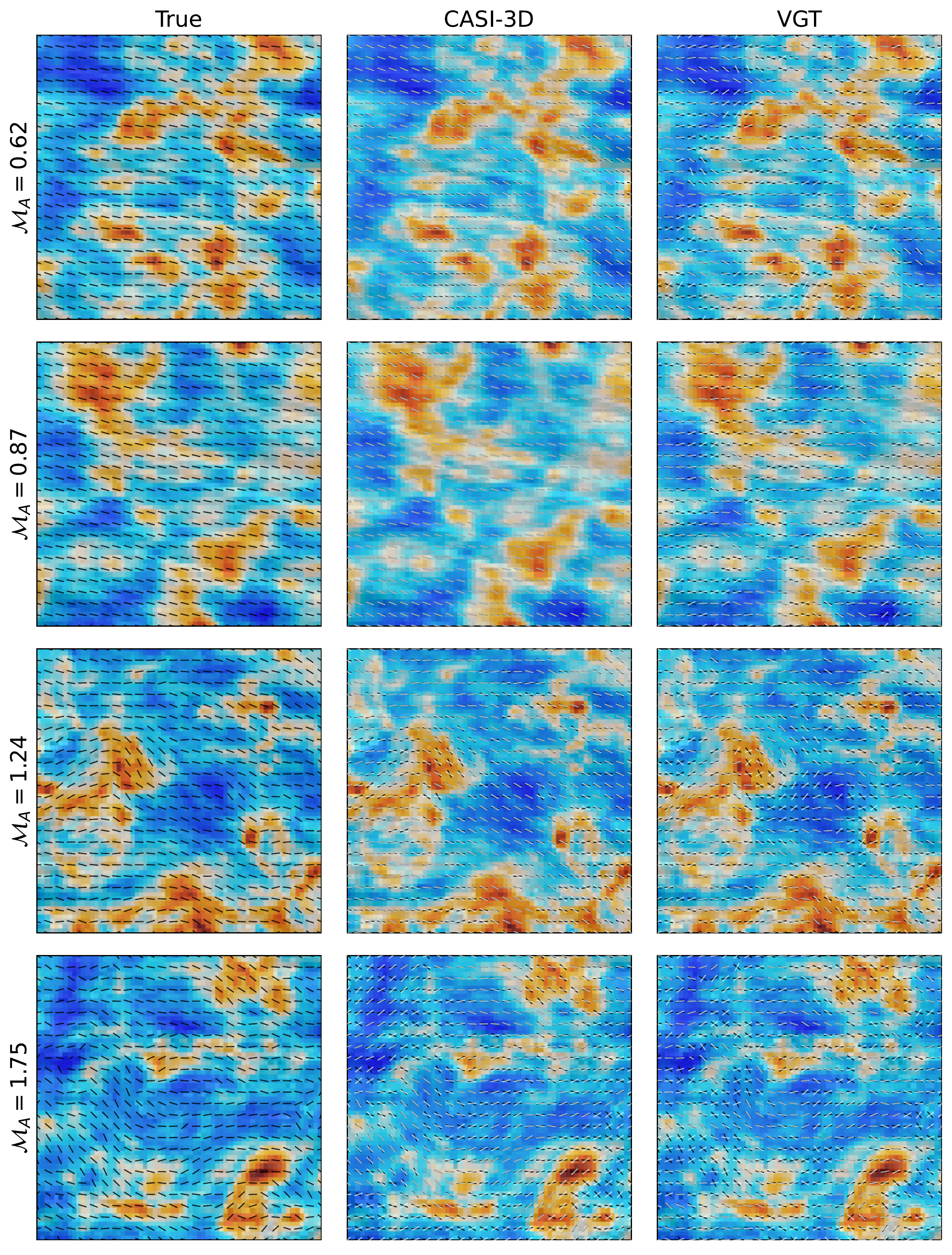}
\caption{Performance of \CASItD\ and VGT to infer the orientation of magnetic fields of four clouds with different Alfv\'en mach numbers. The background is the integrated \13co\ emission. The black lines indicate the true POS magnetic field directions derived from simulations. The gray lines in the middle column indicate the POS magnetic field directions predicted by \CASItD. The gray lines in the right column indicate the POS magnetic field directions calculated by VGT. }
\label{fig.turb-synthetic-co13-bfield-rot-y-pred}
\end{figure*} 

We evaluate the performance on the test samples with different Alfv\'en mach numbers that are not included in the training set. Figure~\ref{fig.turb-synthetic-co13-bfield-rot-y-pred} shows the performance of \CASItD\ to infer the orientation of magnetic fields on four clouds with different Alfv\'en mach numbers. The ``true'' magnetic field directions are derived by a mass-weighted averaging of the POS magnetic field direction along each LOS. It is worth noting that \CASItD\ predicts the POS magnetic field direction at each velocity channel. We average the \CASItD\ predicted POS magnetic field on each LOS by emission intensity weighting. 

In addition, we evaluate the POS magnetic field direction calculated from the VGT method. We follow the same strategy in \citet{2019A&A...622A.166S} and \citet{2020MNRAS.496.4546H} to calculate the gradient of the intensity of each velocity channel. The magnetic field direction is estimated here as the angle of the vector normal to the gradient vector, i.e., $\Phi_{\rm B,VGT} =\Phi_{\rm Gradient} +\frac{\pi}{2}$. Similarly, we average the VGT predicted POS magnetic field on each LOS by emission intensity weighting. As discussed in \citet{2022ApJ...928..132L}, the VGT inferred POS magnetic field direction might be significantly different depending on the choice of velocity range. To make fair comparison in this work, we do not choose specific velocity ranges, but include all channels for the result from VGT and \CASItD. 

As shown in Figure~\ref{fig.turb-synthetic-co13-bfield-rot-y-pred}, \CASItD\ is able to predict the magnetic field orientation with high accuracy in sub-Alfv\'enic clouds. When the magnetic field becomes weaker, the cloud transitions to being trans-Alfv\'enic, where turbulence plays a more significant role in modifying the morphology of the cloud as well as the magnetic field lines. Although \CASItD\ is still able to correctly infer roughly half of the magnetic field direction in trans-Alfv\'enic clouds, it fails at some places where magnetic fields are not the major factor regulating the morphology of the cloud. On the other hand, VGT predicts the magnetic field directions with a lower accuracy, even in a sub-Alfv\'enic clouds. VGT is likely sensitive to the local intensity fluctuations, which might not truly catch the larger scale morphology. This is the reason why VGT requires large block smoothing on the intensity before calculating, which yields a much lower resolution than that from machine learning approaches.

%jct - again, i would change red to white or gray.
\begin{figure*}[hbt!]
\centering
\includegraphics[width=0.68\linewidth]{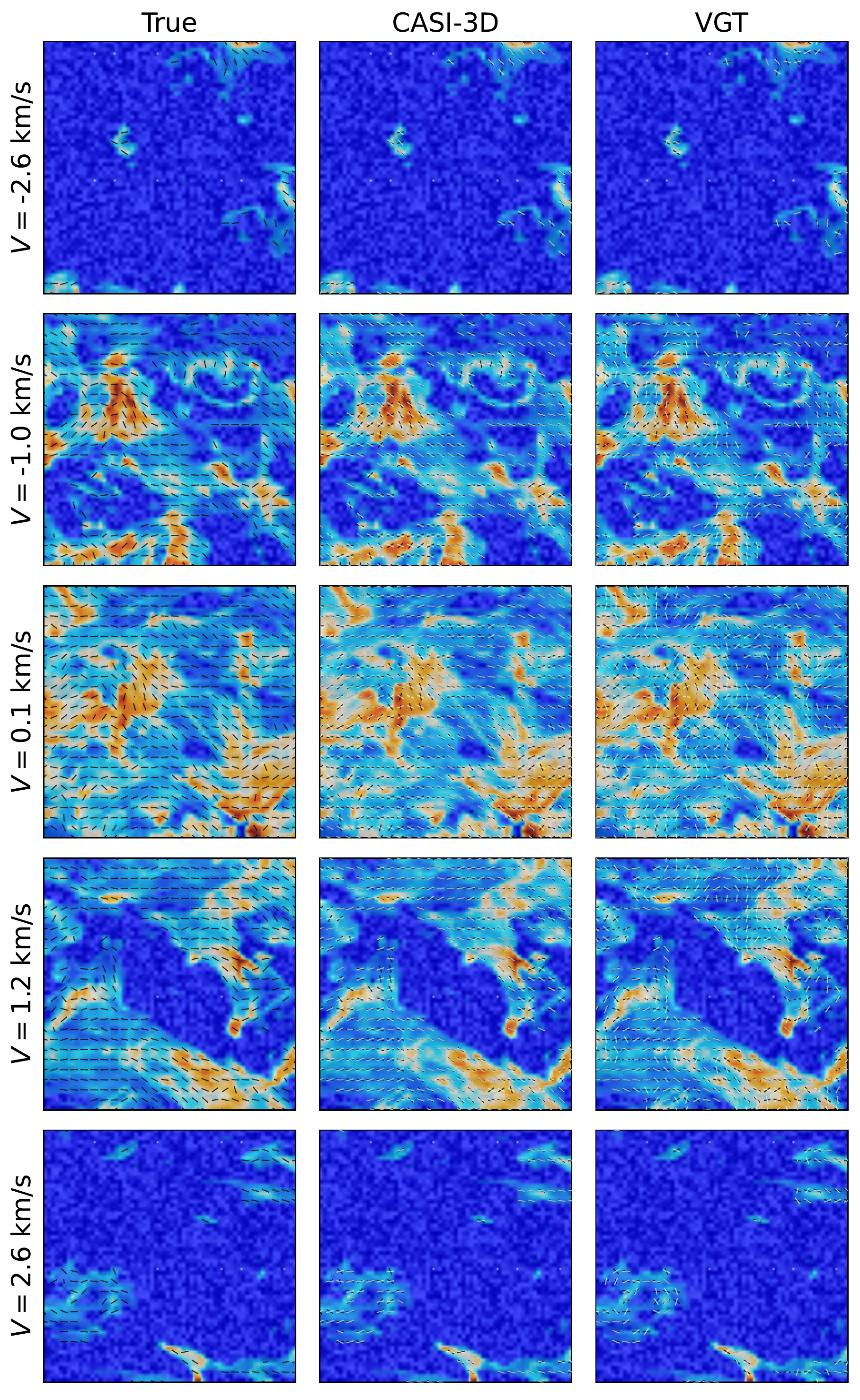}
\caption{ Performance of \CASItD\ and VGT to infer the orientation of magnetic fields across multiple velocity channels on a trans-Alfv\'enic cloud with $\mathcal{M}_{A}=1.24$. The background is the \13co\ emission at each velocity channel. The black lines indicate the true POS magnetic field directions at each velocity channel. The gray lines in the middle column indicate the POS magnetic field directions predicted by \CASItD. The gray lines in the right column indicate the POS magnetic field directions calculated by VGT.}
\label{fig.turb1-synthetic-co13-bfield-rot-y-pred-channel}
\end{figure*}

To better visualize the prediction of a full PPV cube, we show the channel by channel prediction by \CASItD\ on a trans-Alfv\'enic cloud with $\mathcal{M}_{A}=1.24$ in Figure~\ref{fig.turb1-synthetic-co13-bfield-rot-y-pred-channel}. It is obvious that even at the same location the magnetic field direction can be different at different velocities. Neither \CASItD\ nor VGT is able to perfectly infer the magnetic field direction at each velocity. However, it is noticeable that \CASItD\ prediction has more agreement with the true direction than that of VGT, indicating better performance of \CASItD\ than VGT.

\begin{figure*}[hbt!]
\centering
\includegraphics[width=0.88\linewidth]{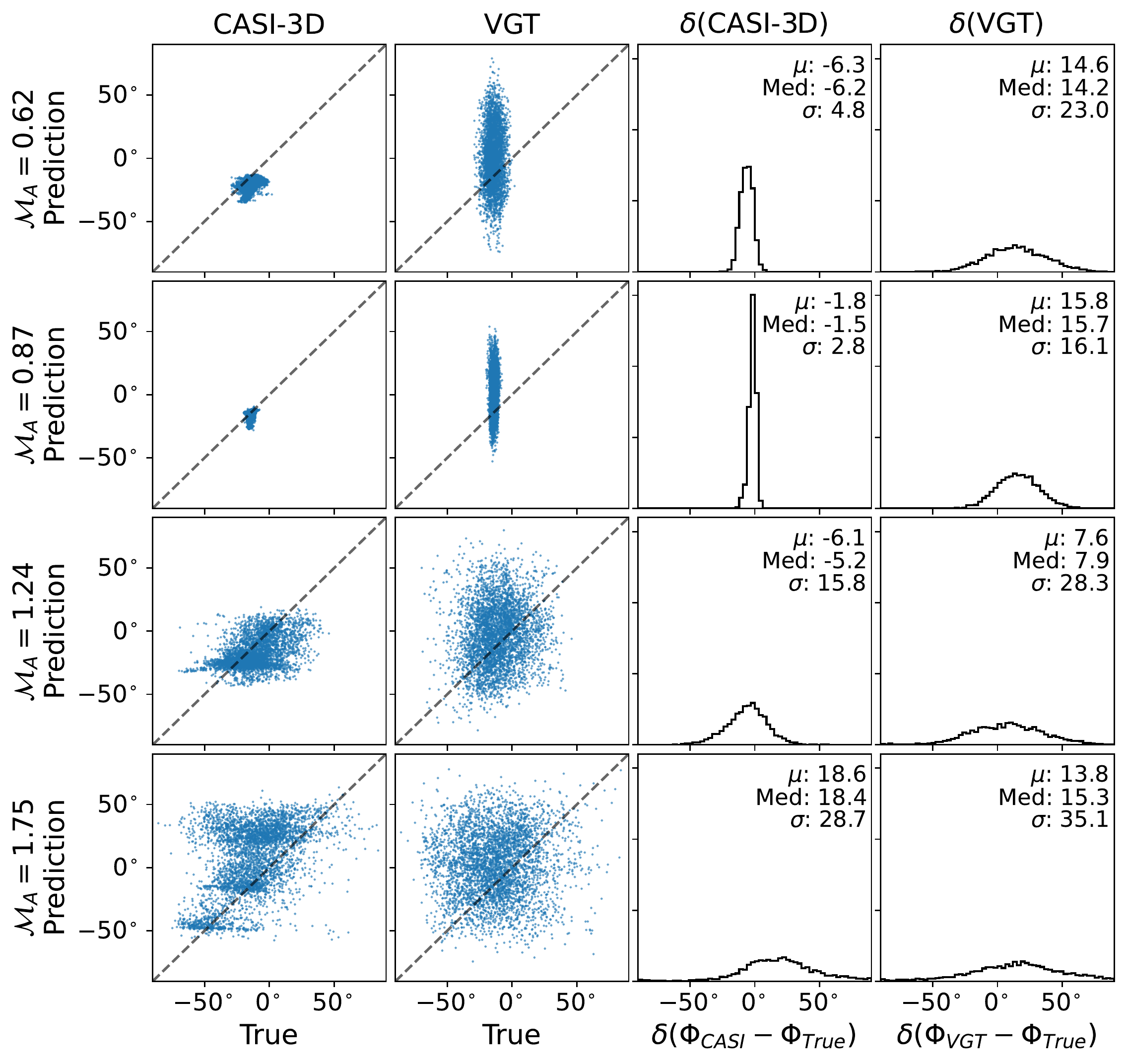}
\caption{First column: scatter plots between the true magnetic field directions $\Phi_{\rm True}$ and the \CASItD\ predicted directions $\Phi_{\rm CASI}$ for the four clouds with different Alfv\'en mach numbers, as labelled in each row. Second column: scatter plots between the true magnetic field directions $\Phi_{\rm True}$ and the VGT predicted directions $\Phi_{\rm VGT}$. Third column: histogram of the angle difference between $\Phi_{\rm True}$ and $\Phi_{\rm CASI}$. Fourth column: histogram of the angle difference between $\Phi_{\rm True}$ and $\Phi_{\rm VGT}$. }
\label{fig.scatte-hist-synthetic-y-rot}
\end{figure*} 

To quantify the performances of \CASItD\ and VGT, in Figure~\ref{fig.scatte-hist-synthetic-y-rot} we present scatter plots between the true magnetic field directions ($\Phi_{\rm True}$) and the predicted ones by \CASItD\ ($\Phi_{\rm CASI}$) and by VGT ($\Phi_{\rm VGT}$).The scatter between $\Phi_{\rm True}$ and $\Phi_{\rm CASI}$ is smaller and closer to the one-to-one line than that between $\Phi_{\rm True}$ and $\Phi_{\rm VGT}$. We also show the histogram of the angle difference between $\Phi_{\rm True}$ and $\Phi_{\rm CASI}$, denoted as $\delta_{\rm CASI-3D}$, and also the histogram of the angle difference between $\Phi_{\rm True}$ and $\Phi_{\rm VGT}$, denoted as $\delta_{\rm VGT}$. It is obvious that $\delta_{\rm CASI-3D}$ has a smaller dispersion than $\delta_{\rm VGT}$ for all different Alfv\'en mach numbers. The dispersion of $\delta_{\rm CASI-3D}$ is the smallest for sub-Alfv\'enic regions, but becomes larger for trans-Alfv\'enic regions. We summarize the statistical results in Table~\ref{tab-stat-casi-vgt-synth-obs-all}.

\begin{table*}[]
\begin{center}
  \caption{Statistical Results of the Performance of \CASItD\ and VGT \label{tab-stat-casi-vgt-synth-obs-all}}
 \begin{tabular}{cccccccc}
\hline
\hline
\multirow{2}{*}{Model} & \multirow{2}{*}{$\mathcal{M}_{A}$}  &  \multicolumn{3}{c}{$\delta_{\rm CASI-3D}$ ($^{\circ}$)} &  \multicolumn{3}{c}{$\delta_{\rm VGT}$ ($^{\circ}$)} \\
  &  &  Mean & Median & Dispersion & Mean & Median & Dispersion \\
\hline
Turb1 & 1.24 & -6.1 &  -5.2 & 15.8 & 7.6 & 7.9 & 28.3 \\ 
Turb2 & 0.62 & -6.3 &  -6.2 & 4.8 & 14.6 & 14.2 & 23.0 \\
Turb3 & 1.75 &  18.6 & 18.4  &  28.7 & 13.8 & 15.3 & 35.1 \\
Turb4 & 0.87 & -1.8 & -1.5 & 2.8 & 15.8 & 15.7 & 16.1 \\
\cline{1-8}
\end{tabular}%
\end{center}
\end{table*}%

%jct - I think we need a quantification of the dispersions of the \delta distributions.

We further examine the performance of \CASItD\ on synthetic observations of other simulations \citep{2003MNRAS.345..325C,2009ApJ...693..250B} run by a different code \citep{2002PhRvL..88x5001C} in Appendix~\ref{Examnine CASI on ENO Data}. We find that \CASItD\ is able to robustly predict magnetic field directions across different sub-Alfv\'enic simulations with low uncertainties ($\lesssim 15 ^{\circ}$).

%jct - some quick summary of the results of this extra comparison should be described here?

\subsection{ \CASItD\ Performance on a Trans-Alfv\'enic Region in Taurus}
\label{CASItD Performance on a Trans-Alfvenic Region in Taurus}

%jct - again I would try white or grey, to see if it is clearer than red.
\begin{figure*}[hbt!]
\centering
\includegraphics[width=0.88\linewidth]{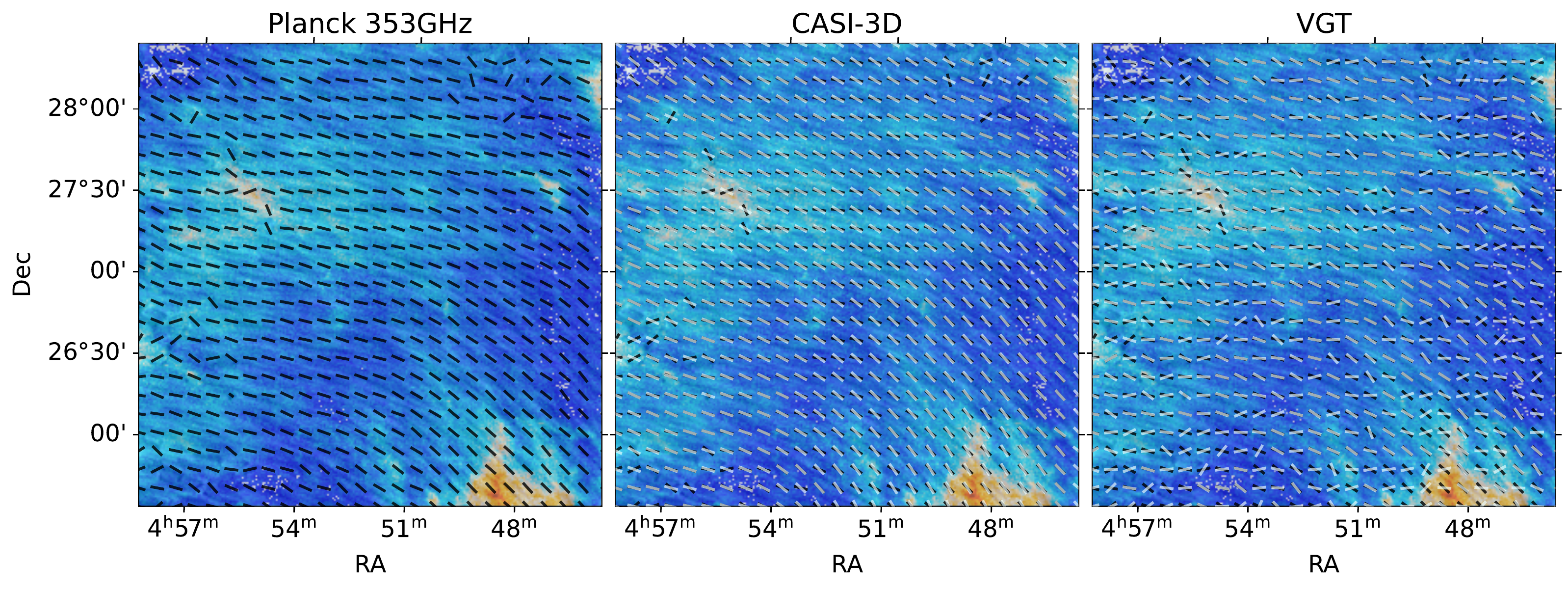}
\caption{Performance of \CASItD\ and VGT to infer the orientation of magnetic fields of the sub-Alfv\'enic region in Taurus. The background is the integrated \co\ emission. The black lines indicate the POS magnetic field directions calculated from {\it Planck} dust polarized emission. The gray lines in the middle column indicate the POS magnetic field directions predicted by \CASItD. The gray lines in the right column indicate the POS magnetic field directions calculated by VGT.}
\label{fig.pred-taurus-tf4-3pannel}
\end{figure*}

In this section, we evaluate the \CASItD\ performance in the analysis of observational data by comparing the \CASItD\ prediction with the dust polarization predicted magnetic field direction. As we discussed above, our \CASItD\ model is mainly trained on sub- and trans-Alfv\'enic clouds. Consequently, we aim to find a sub- and trans-Alfv\'enic region with molecular line emission data as our test sample. The Taurus striations, which were first presented by \citet{2008ApJ...680..428G} and discussed by \citet{2016MNRAS.461.3918H}, are likely sub-Alfv\'enic. \citet{2016MNRAS.461.3918H} proposed that the striations are caused by either the Kelvin–Helmholtz instability or magnetosonic waves propagating, both of which are common in sub-Alfv\'enic clouds.

Figure~\ref{fig.pred-taurus-tf4-3pannel} shows the \co\ integrated intensity of the Taurus striations. We downsample the \co\ data by a factor of 3 to match the pixel resolution of {\it Planck}. We also show the magnetic field orientation inferred from {\it Planck} dust polarization in Figure~\ref{fig.pred-taurus-tf4-3pannel}. We apply the \CASItD\ model to the \co\ cube of this sub-Alfv\'enic region as described in Section~\ref{co and 13co data}. We show the integrated \CASItD\ prediction that is averaged over each line of sight by emission weighting in Figure~\ref{fig.pred-taurus-tf4-3pannel}. We also show the predicted magnetic field directions by VGT in Figure~\ref{fig.pred-taurus-tf4-3pannel}. 

It is clear that the \CASItD\ prediction agrees with the dust polarization inferred magnetic direction at most locations. There are some localized regions of misalignment where the dust polarization inferred magnetic direction changes significantly. \CASItD\ is not sensitive to these small scale fluctuations. \CASItD\ predicts the magnetic field direction of each pixel by learning the morphology from it and its surroundings. Consequently, \CASItD\ tends to predict a smoother magnetic field direction without sharp fluctuations, which is a limitation of this machine learning method. On the other hand, the prediction by VGT has a larger deviation from the dust polarization inferred magnetic direction.

\begin{figure*}[hbt!]
\centering
\includegraphics[width=0.58\linewidth]{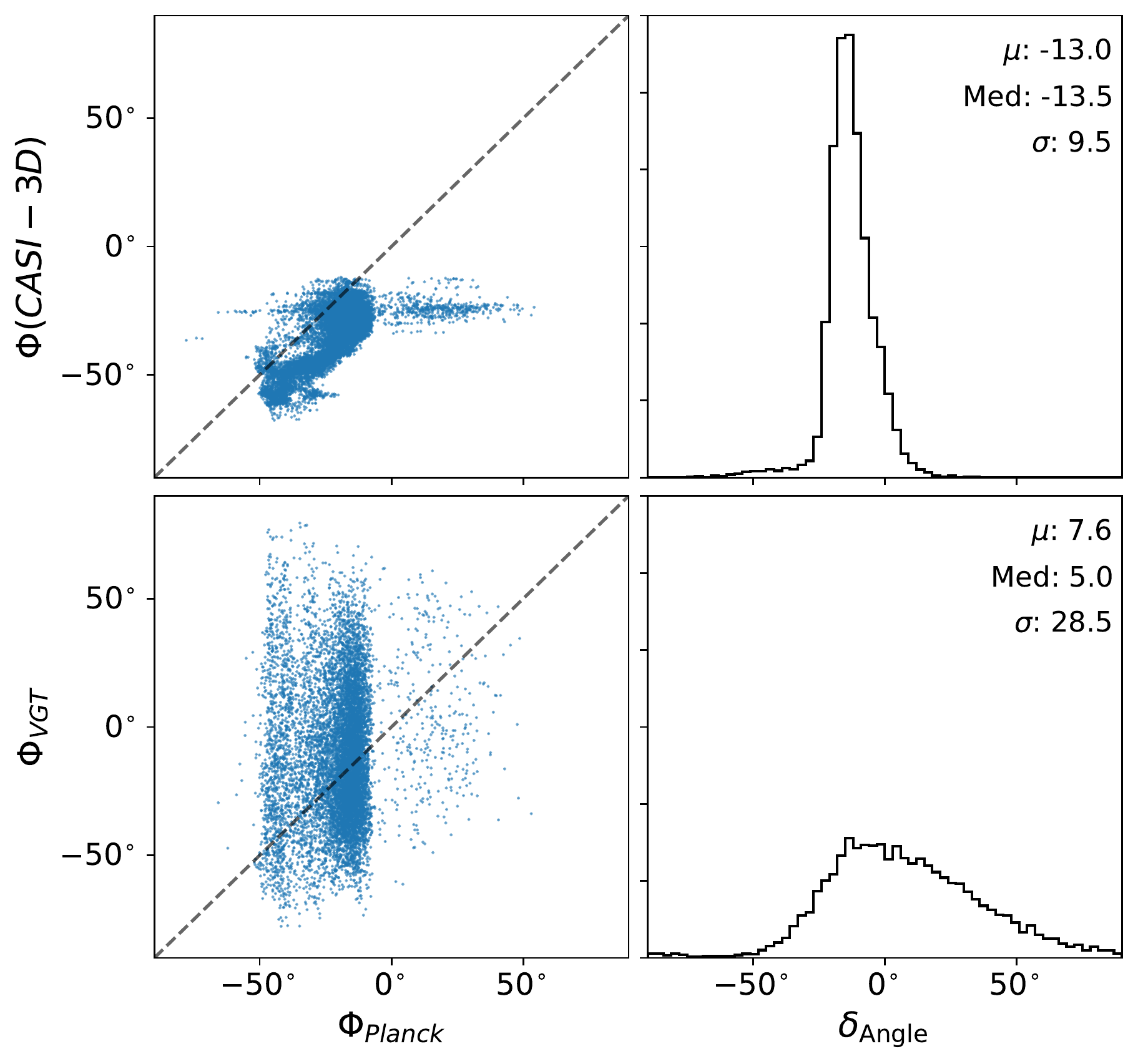}
\caption{Upper left: scatter plot between the {\it Planck} dust polarization inferred magnetic field directions $\Phi_{\rm Planck}$ and the \CASItD\ predicted directions $\Phi_{\rm CASI}$ for the sub-Alfv\'enic region in Taurus. Lower left: scatter plot between the {\it Planck} dust polarization inferred magnetic field directions $\Phi_{\rm Planck}$ and the VGT predicted directions $\Phi_{\rm VGT}$. Upper right: histogram of the angle difference between $\Phi_{\rm Planck}$ and $\Phi_{\rm CASI}$. Lower right: histogram of the angle difference between $\Phi_{\rm Planck}$ and $\Phi_{\rm VGT}$. }
\label{fig.scatter-taurus-tf4}
\end{figure*}

To quantify the performance of \CASItD\ and that of VGT, in Figure~\ref{fig.pred-taurus-tf4-3pannel} we present scatter plots between the dust polarization inferred magnetic direction $\Phi_{\rm Planck}$ and the predicted ones by \CASItD\ $\Phi_{\rm CASI}$, and by VGT $\Phi_{\rm VGT}$. It is clear that the scatter of between $\Phi_{\rm Planck}$ and $\Phi_{\rm CASI}$ is closer to the one-to-one line than that between $\Phi_{\rm Planck}$ and $\Phi_{\rm VGT}$. 
%jct - we need to quantify the offset and the scatter
We show the histogram of the angle difference between $\Phi_{\rm Planck}$ and $\Phi_{\rm CASI}$, , denoted as $\delta_{\rm CASI-Planck}$, and also the angle difference between $\Phi_{\rm Planck}$ and $\Phi_{\rm VGT}$, denoted as $\delta_{\rm VGT-Planck}$. It is clear that \CASItD\ prediction is closer to the dust polarization inferred magnetic direction. The dispersion of $\delta_{\rm CASI-Planck}$ is 9.5~\deg, which is much smaller than the dispersion of  $\delta_{\rm VGT-Planck}$, which is 28.5\deg. 

{ It is worth noting that there are effects that can lead to $\Phi_{\rm Planck}$ yielding an inaccurate estimate of magnetic field direction in CO-emitting molecular gas due to the contaminating effects of polarized dust emission from atomic regions or ``CO-dark'' molecular regions along the line of sight (LOS).  It has long been known that such regions contain dust. For example, \citet{1992ApJ...397..165J} found spatial correlations between H{\sc i} emission and infrared circus in filamentary clouds. 
%jct - what are "filament regions"?
\citet{2011A&A...536A..24P} carried out a more systematic study on the correlation between dust emission and H{\sc i} emission over 800 deg$^{2}$ at high Galactic latitudes, and found that dust emission at lower column densities ($\leq 2\times 10^{20}$~cm$^{-2}$) is well correlated with H{\sc i} emission. This indicates that in such diffuse regions the hydrogen is predominantly
in the neutral atomic phase where \co\ is absent, but dust still exists. Moreover, it is also known that there are significant amounts of molecular gas that are dark in \co\ emission \citep{2005Sci...307.1292G,2016ApJ...819...22X} due to photodissociation of CO \citep[e.g.,][]{1999RvMP...71..173H} or destruction by cosmic rays \citep[e.g.,][]{2015ApJ...803...37B,2021MNRAS.502.2701B}. \citet{2016ApJ...819...22X} found that the dark molecular gas fraction can be up to 80\% at $A_{V} \sim 1$~mag in local Galactic clouds. 
%jct - check above
Thus, perfect agreement between dust polarization based and \co\ based $B$-field measurements is not expected in general due to LOS dust contamination.
%and  \co\ data can not exactly show the entire dust emitting region.} There { must} be LOS dust emission, e.g., foreground and background that are not from the \co\ emitting location. This LOS contamination { will definitely} influence the dust polarization direction. 
This effect could help explain the systematic offset of about -13\deg\ that is seen between $\Phi_{\rm CASI}$ and $\Phi_{\rm Planck}$}.

\subsection{ \CASItD\ Performance on the Full Taurus \co\ and \13co\ Map}
\label{CASItD Performance on the Full Taurus Map}

\begin{figure*}[hbt!]
\centering
\includegraphics[width=0.98\linewidth]{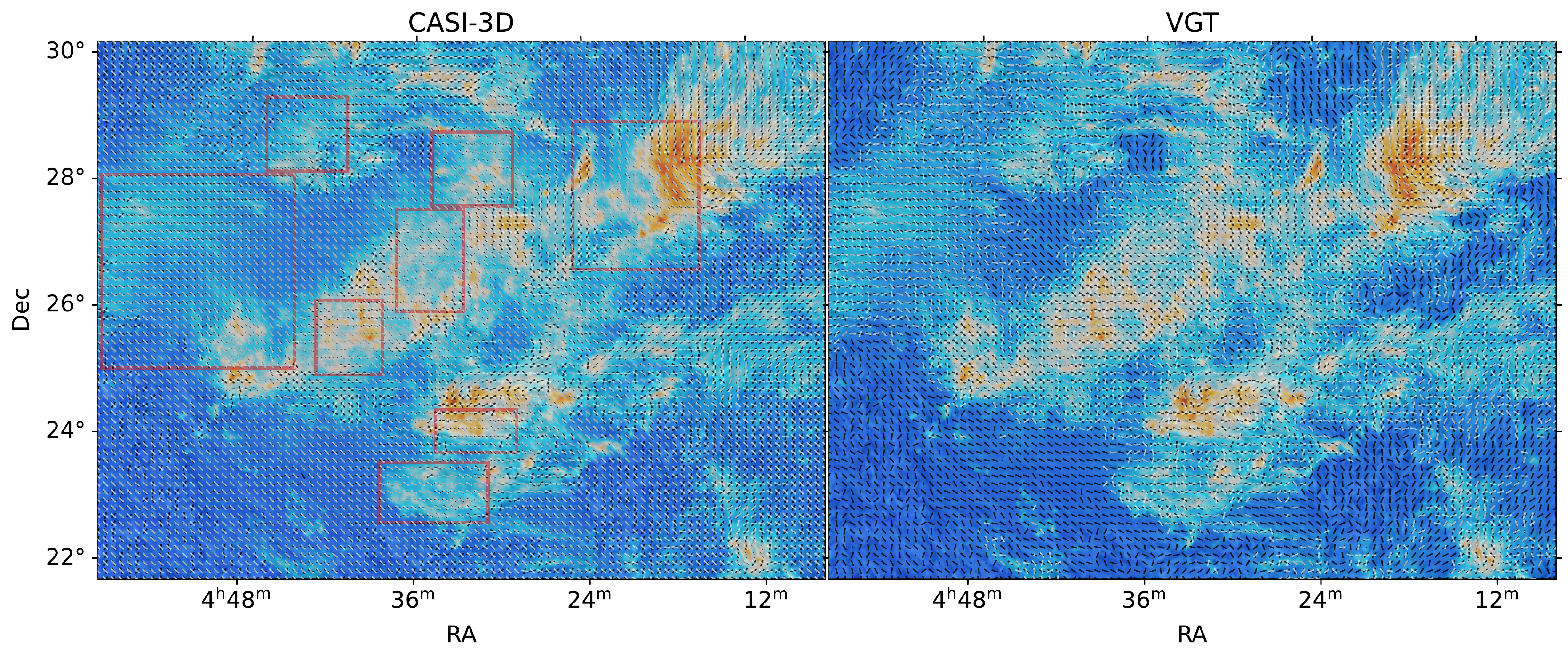}
\caption{Performance of \CASItD\ and VGT to infer the orientation of magnetic fields of the full Taurus map. The background is the integrated \co\ emission. The black lines indicate the POS magnetic field directions calculated from {\it Planck} dust polarized emission. The gray lines in the left panel indicate the POS magnetic field directions predicted by \CASItD. The gray lines in the right panel indicate the POS magnetic field directions calculated by VGT. The red boxes highlight some regions, where \CASItD\ inferred magnetic field direction is similar to that by {\it Planck} dust polarized emission.}
\label{fig.pred-taurus-tf2-all-2panel-12co}
\end{figure*}

\begin{figure*}[hbt!]
\centering
\includegraphics[width=0.98\linewidth]{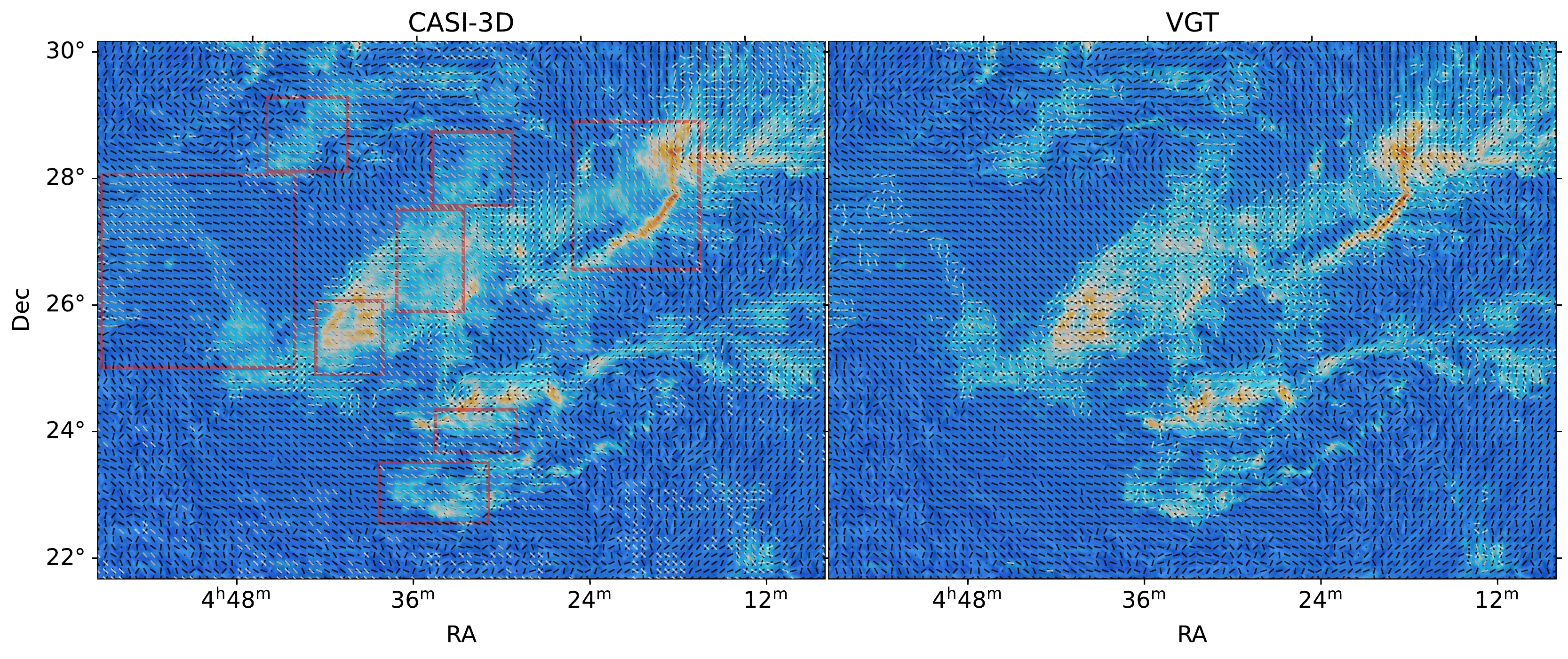}
\caption{Same as Figure~\ref{fig.pred-taurus-tf2-all-2panel-12co}, but for \13co.}
\label{fig.pred-taurus-tf2-all-2panel-13co}
\end{figure*}

In this section, we apply our \CASItD\ model to the full Taurus \co\ and \13co\ map. We compare the \CASItD\ predicted magnetic field directions with the dust polarization predicted magnetic field directions. 

Figure~\ref{fig.pred-taurus-tf2-all-2panel-12co} and \ref{fig.pred-taurus-tf2-all-2panel-13co} show the performance of \CASItD\ to predict magnetic field directions on the full Taurus \co\ and \13co\ emission data. We also show the magnetic field directions inferred by VGT for comparison. Note that our current \CASItD\ model is only trained on sub- and trans- Alfv\'enic turbulent simulations, which do not include self gravity. Consequently, if there is self gravity or other dynamical processes, such as feedback or large-scale converging gas flows, \CASItD\ is not able to correctly infer the direction of magnetic fields. However, if \CASItD\ predicted magnetic field directions are similar to those inferred by {\it Planck} dust polarized emission, it is likely that this region is closer to the trans-Alfvénic regime, indicating relatively strong magnetic field strengths. We highlight some of these regions in red boxes in Figure~\ref{fig.pred-taurus-tf2-all-2panel-12co} and \ref{fig.pred-taurus-tf2-all-2panel-13co}, where \CASItD\ inferred magnetic field direction is very similar to that inferred from {\it Planck} dust polarized emission. 

\begin{figure*}[hbt!]
\centering
\includegraphics[width=0.98\linewidth]{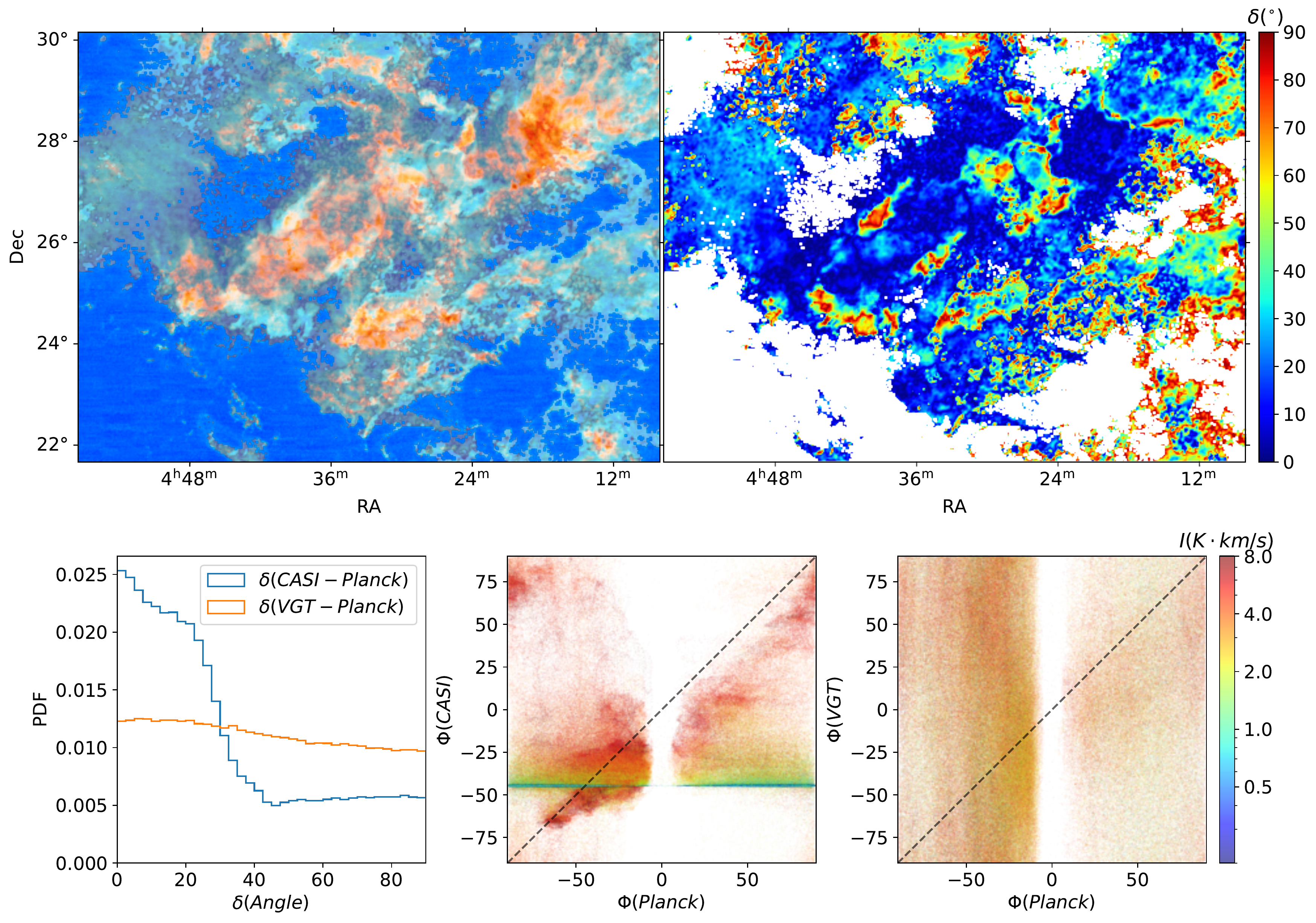}
\caption{Upper left: integrated \co\ emission overlaid with gray shadows indicating the location where \CASItD\ predicted magnetic field directions are similar to those inferred by {\it Planck} dust polarized emission. Upper right: angle difference between the \CASItD\ predicted magnetic field directions and those inferred by {\it Planck} dust polarized emission. Lower left: distribution of the angle difference between the \CASItD\ predicted magnetic field directions and those inferred by {\it Planck}, and the angle difference between VGT inferred magnetic field directions and those inferred by {\it Planck}. Lower middle: scatter plot between the Planck dust polarization inferred magnetic field directions $\Phi_{\rm Planck}$ and the \CASItD\ predicted directions $\Phi_{\rm CASI}$. Lower right: scatter plot between the Planck dust polarization inferred magnetic field directions $\Phi_{\rm Planck}$ and the VGT predicted directions $\Phi_{\rm VGT}$. The color in the scatter plots indicate the integrated intensity of \co\ at each pixel.}
\label{fig.pred-taurus-tf2-all-5panel-scat-12co}
\end{figure*}

\begin{figure*}[hbt!]
\centering
\includegraphics[width=0.98\linewidth]{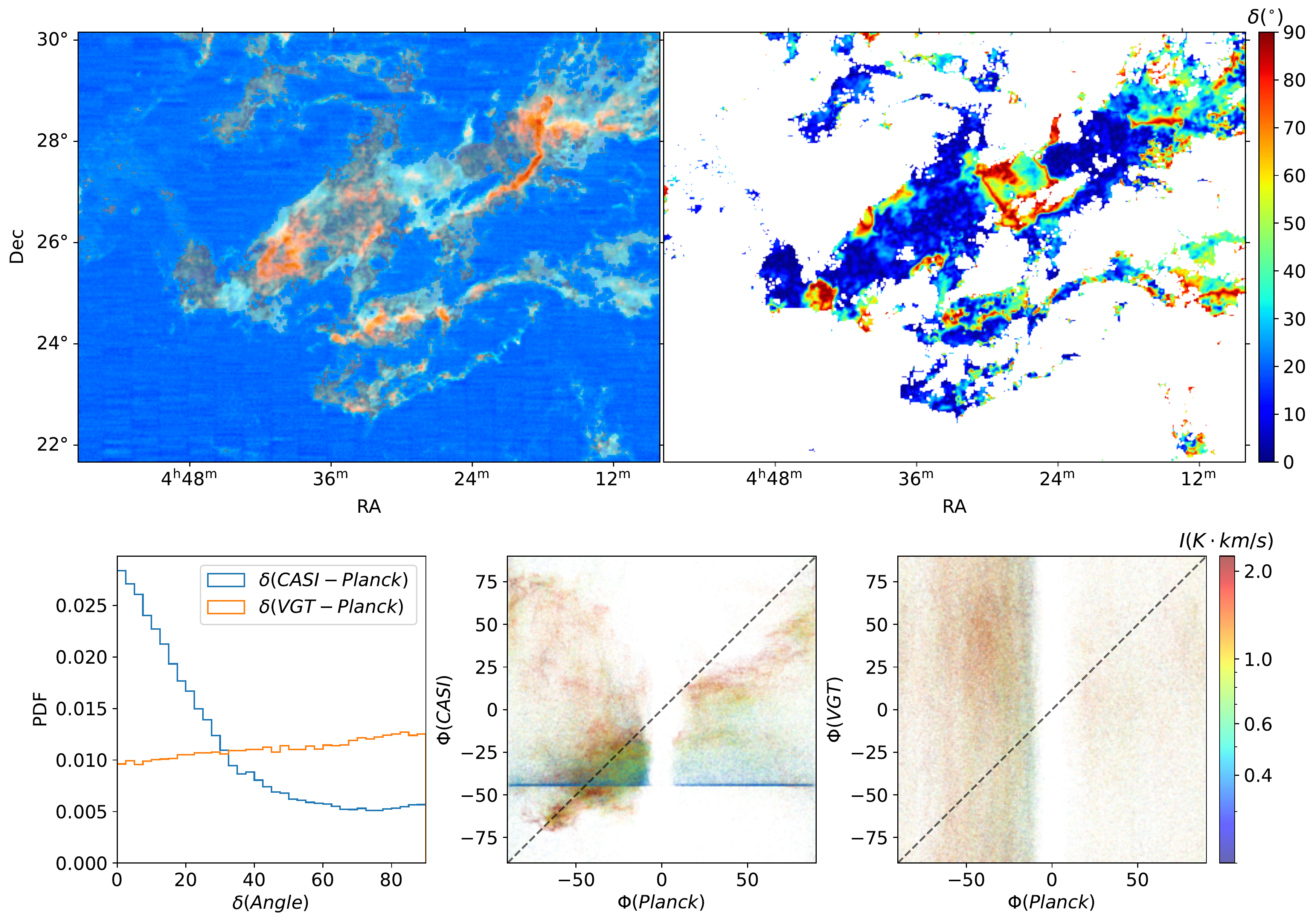}
\caption{Same as Figure~\ref{fig.pred-taurus-tf2-all-5panel-scat-12co}, but for \13co.}
\label{fig.pred-taurus-tf2-all-5panel-scat-13co}
\end{figure*}

To better evaluate the performance of \CASItD\, we show the angle difference between the \CASItD\ predicted magnetic field directions and those inferred by {\it Planck} dust polarized emission in Figures~\ref{fig.pred-taurus-tf2-all-5panel-scat-12co} and \ref{fig.pred-taurus-tf2-all-5panel-scat-13co}. In most regions, the angle difference between the \CASItD\ predicted magnetic field directions and those inferred by {\it Planck} dust polarized emission, denoted as $\delta(\rm CASI-Planck)$, is less than 20\deg, as shown in the shadowed region in Figures~\ref{fig.pred-taurus-tf2-all-5panel-scat-12co} and \ref{fig.pred-taurus-tf2-all-5panel-scat-13co}. This indicates that self gravity is likely not dominant in these regions. On the other hand, in the upper right panel of Figure~\ref{fig.pred-taurus-tf2-all-5panel-scat-12co} and \ref{fig.pred-taurus-tf2-all-5panel-scat-13co}, we notice that in some regions %, especially those with higher column densities, that
the magnetic field directions inferred by \CASItD\ are dissimilar ($\geq 45$\deg), even perpendicular, to those inferred by dust polarization. This is likely caused by self gravity or large-scale gas flows, where the gas morphology is not particularly regulated by magnetic fields \citep[e.g.,][]{2017ApJ...836...95O,2022ApJ...928..132L}.
It is worth noting that { not all high column density regions show large differences} between \CASItD\ predicted magnetic field directions and those inferred by {\it Planck} dust polarized emission. As shown in the scatter plots in Figures~\ref{fig.pred-taurus-tf2-all-5panel-scat-12co} and \ref{fig.pred-taurus-tf2-all-5panel-scat-13co}, a significant amount of pixels with small $\delta(\rm CASI-Planck)$ also have strong integrated intensities. 

The magnetic field directions inferred by \CASItD\ on \co\ and \13co\ are mostly consistent, especially in the red boxes in Figure~\ref{fig.pred-taurus-tf2-all-2panel-12co} and \ref{fig.pred-taurus-tf2-all-2panel-13co}. This demonstrates the robustness of machine learning to handle different molecular emission data. Although \co\ is optically thick near the central velocity channels, \CASItD\ is still able to capture their morphology to infer the magnetic field directions \citep[however, for discussion of the effect of optical depth, see, e.g.][]{2020MNRAS.496.4546H,2022ApJ...928..132L}. When comparing the magnetic field directions inferred by VGT and that by {\it Planck} dust polarized emission, we notice significant fluctuations on small scales. This in turn proves that VGT likely requires large sub-block averaging to obtain a robust result. Consequently, it is almost impossible to derive a high-resolution magnetic field map by VGT, but possible by \CASItD. The current Taurus \co\ and \13co\ observation has a pixel resolution of 23\arcsec, which is three times higher than that of the {\it Planck} dust emission map (1$^{\prime}$.07.). In the future, we will apply the \CASItD\ model to the FUGIN (FOREST Unbiased Galactic plane Imaging survey with the Nobeyama 45-m telescope) project \citep{2017PASJ...69...78U}, which will yield maps that are a factor of nine higher resolution than those from {\it Planck}.

\section{Conclusions}
\label{Conclusions}

We have trained the deep learning method \CASItD\ to predict the orientation of magnetic fields in sub- and trans- Alfv\'enic turbulent clouds from molecular line emission. We have tested the \CASItD\ performance on synthetic test samples and real observational data. Our main findings are as follows:

\begin{enumerate}

\item  \CASItD\ is able to predict the magnetic field directions from molecular line emission at a pixel level with higher accuracy than that from VGT for sub- and trans-Alfv\'enic clouds. 
%jct - quote quantitative results

\item \CASItD\ achieves higher accuracy in sub-Alfv\'enic clouds ($\lesssim 10^{\circ}$) than that in trans-Alfv\'enic clouds ($\lesssim 30^{\circ}$). %jct - quote quantitative results
This is consistent with our intuition, where stronger magnetic fields play a more important role in regulating the morphology of clouds. 

\item \CASItD\ is able to infer the magnetic field direction of a sub-Alfv\'enic region in Taurus from \co\ emission. The prediction is consistent with the {\it Planck} dust polarized emission inferred magnetic field directions, which has a systematical offset of -13\deg\ and a dispersion of 9.5\deg.
%jct quote quantitative results.

\item We have applied \CASItD\ to the full Taurus \co\ and \13co\ maps, yielding $B-$field morphology information at a three times higher angular resolution than the {\it Planck}-derived map. We find that in many regions \CASItD\ inferred magnetic field directions are similar to those inferred from {\it Planck} dust polarized emission. This implies that much of the Taurus region is in the sub-Alfvénic regime, i.e., subject to relatively strong, dynamically important magnetic fields.

\end{enumerate}

{ We thank the anonymous referee for comments that improved this manuscript.} D.X. acknowledges support from the Virginia Initiative on Cosmic Origins (VICO). C-Y.L acknowledges support from an ESO Ph.D. studentship. J.C.T. acknowledges support from NSF grant AST-2009674 and ERC Advanced Grant MSTAR. D.X. thanks Zhi-Yun Li for helpful discussions. The authors acknowledge Research Computing at The University of Virginia for providing computational resources and technical support that have contributed to the results reported within this publication.

\appendix
\section{Examination of \CASItD\ Performance on Different Simulations}
\label{Examnine CASI on ENO Data}
In this section, we evaluate the performance of \CASItD\ on the synthetic observations of the simulations from \citet{2003MNRAS.345..325C} and \citet{2009ApJ...693..250B}. The simulations adopt a second-order-accurate hybrid essentially non-oscillatory (ENO) scheme \citep{2002PhRvL..88x5001C} to solve the ideal MHD equations in a periodic
box. The turbulence is driven solenoidally at wave scale $k$ equal to about 2.5, which is different to our simulation in Section~\ref{Magnetohydrodynamics Simulations}. In our simulations, we drive turbulence with $1\le k \le 2$, and $\frac{2}{3}$ of the total power is in solenoidal motions and the rest $\frac{1}{3}$ is in compressive motions. This indicates no bias of imposing solenoidal or compressive modes \citep{1995ApJ...448..226D}. We adopt the run with $\mathcal{M}_{S}=7$ and $\mathcal{M}_{A}=0.7$. We assume a kinetic temperature of 20 K, which yields a sound speed of 0.27 \kms\ and a turbulent velocity of 1.86 \kms. We assume a mean density of 523~\cmc, which is the same as our runs. This indicates a mean magnetic field strength of 12 $\mu G$. We conduct radiative transfer following exactly the same process as Section~\ref{Training Sets} to generate \co\ and \13co\ emission. 

\begin{figure*}[hbt!]
\centering
\includegraphics[width=0.88\linewidth]{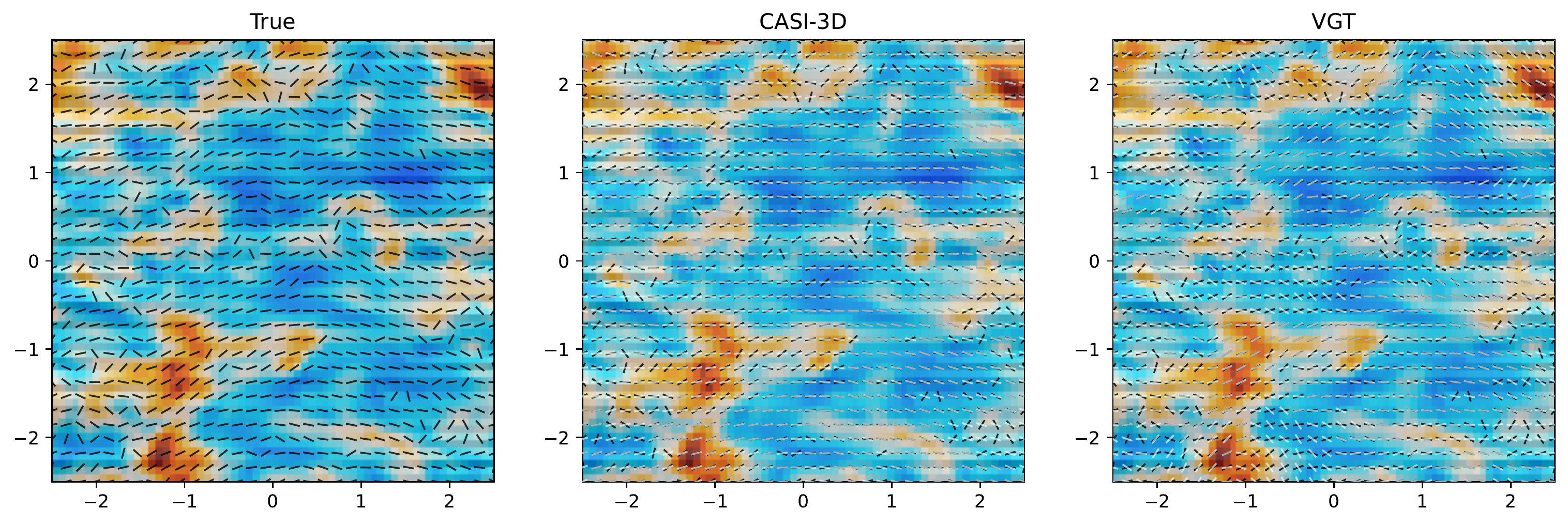}
\caption{Performance of \CASItD\ and VGT to infer the orientation of magnetic fields on the synthetic \co\ data of ENO simulations.}
\label{fig.pred-ENO-3pannel}
\end{figure*}

\begin{figure*}[hbt!]
\centering
\includegraphics[width=0.58\linewidth]{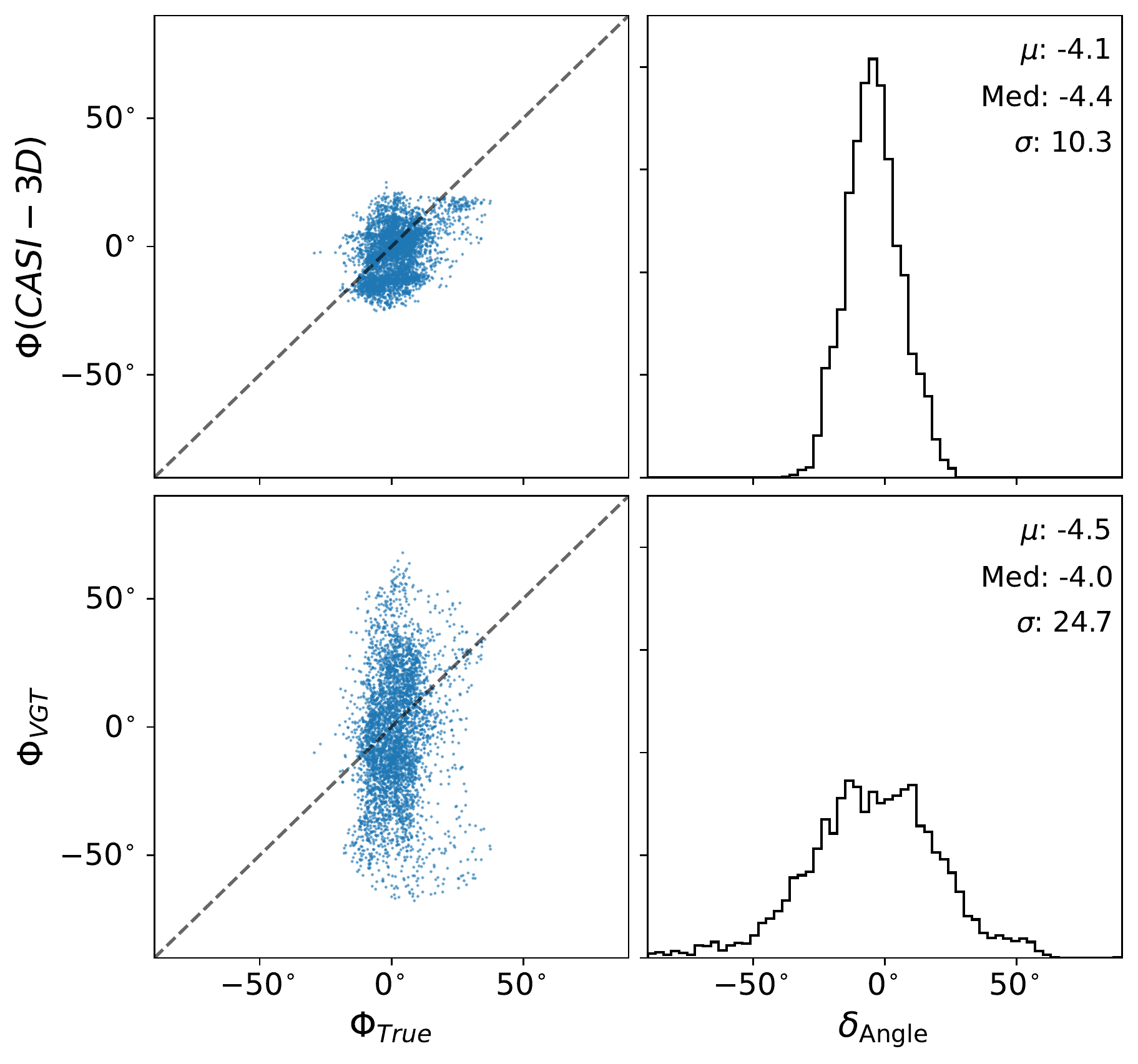}
\caption{Upper left: scatter plot between the true magnetic field directions $\Phi_{\rm True}$ and the \CASItD\ predicted directions $\Phi_{\rm CASI}$ for the synthetic \co\ data of ENO simulations. Lower left: scatter plot between the true magnetic field directions $\Phi_{\rm True}$ and the VGT predicted directions $\Phi_{\rm VGT}$. Upper right: the histogram of the angle difference between $\Phi_{\rm Planck}$ and $\Phi_{\rm CASI}$. Lower right: the histogram of the angle difference between $\Phi_{\rm Planck}$ and $\Phi_{\rm VGT}$. }
\label{fig.scatter-ENO-1}
\end{figure*}

\begin{figure*}[hbt!]
\centering
\includegraphics[width=0.68\linewidth]{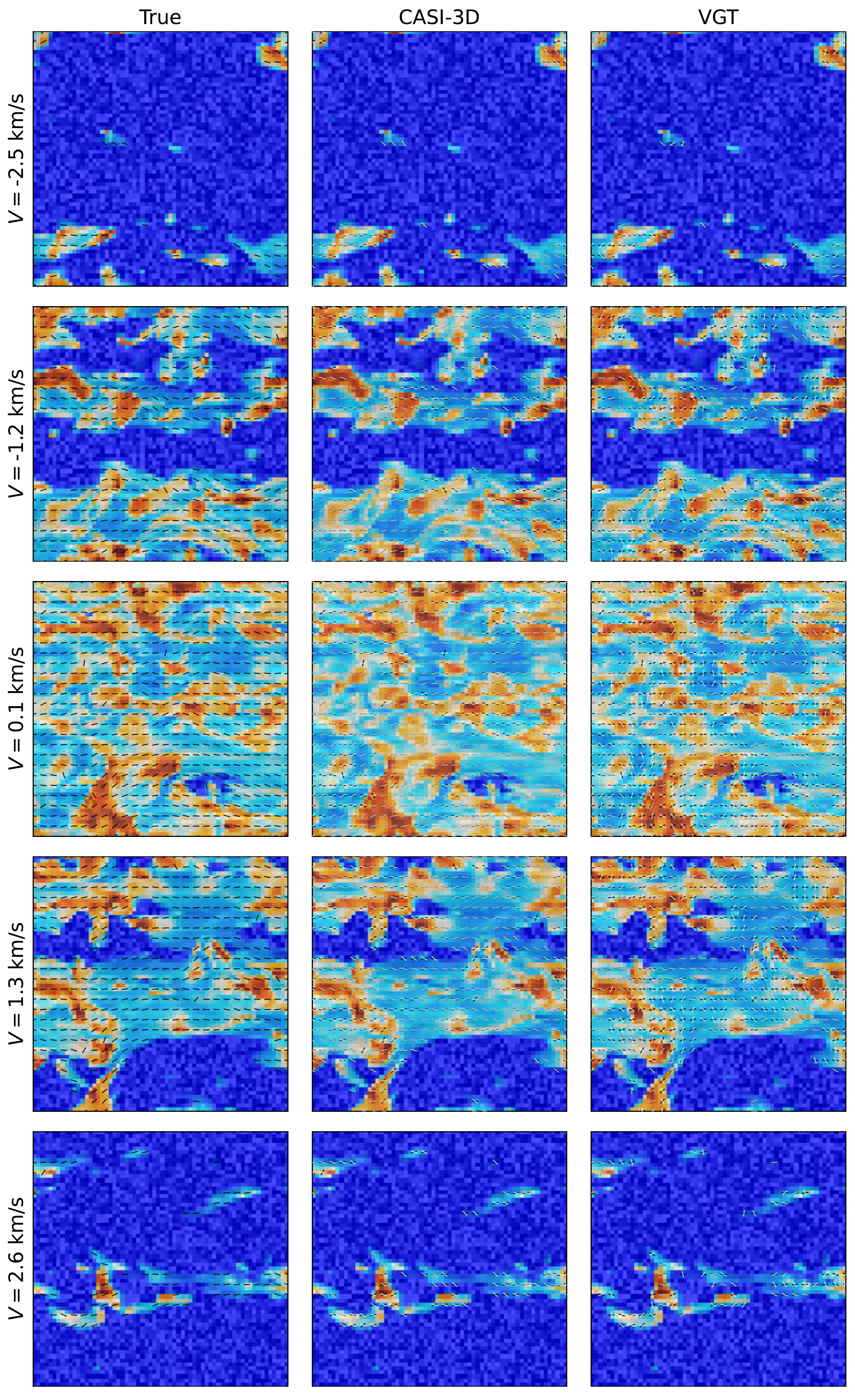}
\caption{Performance of \CASItD\ and VGT to infer the orientation of magnetic fields across multiple velocity channels on the synthetic \co\ data of ENO simulations. The background is the \co\ emission at each velocity channel. The black lines indicate the true POS magnetic field directions at each velocity channel. The gray lines in the middle column indicate the POS magnetic field directions predicted by \CASItD. The gray lines in the right column indicate the POS magnetic field directions calculated by VGT.}
\label{fig.turb1-synthetic-ENO-channel}
\end{figure*}

Figure~\ref{fig.pred-ENO-3pannel} shows the performance of \CASItD\ to infer the orientation of magnetic fields on the synthetic \co\ data of ENO simulations. We present scatter plots between the true magnetic field directions $\Phi_{\rm True}$ and the predicted ones by \CASItD\ $\Phi_{\rm CASI}$, and by VGT $\Phi_{\rm VGT}$ in Figure~\ref{fig.scatter-ENO-1}. We also show the channel by channel prediction by \CASItD\ in Figure~\ref{fig.turb1-synthetic-ENO-channel}. To sum up, the dispersion of $\delta_{\rm CASI}$ is 10.3~\deg, which is much smaller than the dispersion of  $\delta_{\rm VGT}$, which is 24.7\deg\ on the synthetic ENO data sets. \CASItD\ performs robustly in predicting magnetic field directions across different simulations, which provides confidence to apply \CASItD\ to other observational data. 
%jct - quote some quantitative results on the dispersions

%jct - try red to white or gray?

\bibliographystyle{aasjournal}
\bibliography{references}

\begin{thebibliography}{}
\expandafter\ifx\csname natexlab\endcsname\relax\def\natexlab#1{#1}\fi
\providecommand{\url}[1]{\href{#1}{#1}}
\providecommand{\dodoi}[1]{doi:~\href{http://doi.org/#1}{\nolinkurl{#1}}}
\providecommand{\doeprint}[1]{\href{http://ascl.net/#1}{\nolinkurl{http://ascl.net/#1}}}
\providecommand{\doarXiv}[1]{\href{https://arxiv.org/abs/#1}{\nolinkurl{https://arxiv.org/abs/#1}}}

\bibitem[{{Bai} {et~al.}(2021){Bai}, {Liu}, {Deng}, {Jiang}, {Guo}, {Bi},
  {Feng}, {Jin}, {Cao}, {Su}, \& {Ji}}]{2021A&A...652A.143B}
{Bai}, X., {Liu}, H., {Deng}, Y., {et~al.} 2021, \aap, 652, A143,
  \dodoi{10.1051/0004-6361/202140374}

\bibitem[{{Beck} \& {Graeve}(1982)}]{1982A&A...105..192B}
{Beck}, R., \& {Graeve}, R. 1982, \aap, 105, 192

\bibitem[{{Bisbas} {et~al.}(2015){Bisbas}, {Papadopoulos}, \&
  {Viti}}]{2015ApJ...803...37B}
{Bisbas}, T.~G., {Papadopoulos}, P.~P., \& {Viti}, S. 2015, \apj, 803, 37,
  \dodoi{10.1088/0004-637X/803/1/37}

\bibitem[{{Bisbas} {et~al.}(2021){Bisbas}, {Tan}, \&
  {Tanaka}}]{2021MNRAS.502.2701B}
{Bisbas}, T.~G., {Tan}, J.~C., \& {Tanaka}, K. E.~I. 2021, \mnras, 502, 2701,
  \dodoi{10.1093/mnras/stab121}

\bibitem[{{Burkhart} {et~al.}(2009){Burkhart}, {Falceta-Gon{{c}}alves},
  {Kowal}, \& {Lazarian}}]{2009ApJ...693..250B}
{Burkhart}, B., {Falceta-Gon{{c}}alves}, D., {Kowal}, G., \& {Lazarian}, A.
  2009, \apj, 693, 250, \dodoi{10.1088/0004-637X/693/1/250}

\bibitem[{{Burn}(1966)}]{1966MNRAS.133...67B}
{Burn}, B.~J. 1966, \mnras, 133, 67, \dodoi{10.1093/mnras/133.1.67}

\bibitem[{{Cho} \& {Lazarian}(2002)}]{2002PhRvL..88x5001C}
{Cho}, J., \& {Lazarian}, A. 2002, \prl, 88, 245001,
  \dodoi{10.1103/PhysRevLett.88.245001}

\bibitem[{{Cho} \& {Lazarian}(2003)}]{2003MNRAS.345..325C}
---. 2003, \mnras, 345, 325, \dodoi{10.1046/j.1365-8711.2003.06941.x}

\bibitem[{{Clark} {et~al.}(2015){Clark}, {Hill}, {Peek}, {Putman}, \&
  {Babler}}]{2015PhRvL.115x1302C}
{Clark}, S.~E., {Hill}, J.~C., {Peek}, J.~E.~G., {Putman}, M.~E., \& {Babler},
  B.~L. 2015, \prl, 115, 241302, \dodoi{10.1103/PhysRevLett.115.241302}

\bibitem[{{Clark} {et~al.}(2019){Clark}, {Peek}, \&
  {Miville-Desch{\^e}nes}}]{2019ApJ...874..171C}
{Clark}, S.~E., {Peek}, J.~E.~G., \& {Miville-Desch{\^e}nes}, M.~A. 2019, \apj,
  874, 171, \dodoi{10.3847/1538-4357/ab0b3b}

\bibitem[{{Clark} {et~al.}(2014){Clark}, {Peek}, \&
  {Putman}}]{2014ApJ...789...82C}
{Clark}, S.~E., {Peek}, J.~E.~G., \& {Putman}, M.~E. 2014, \apj, 789, 82,
  \dodoi{10.1088/0004-637X/789/1/82}

\bibitem[{{Crutcher}(1999)}]{1999ApJ...520..706C}
{Crutcher}, R.~M. 1999, \apj, 520, 706, \dodoi{10.1086/307483}

\bibitem[{{Crutcher}(2012)}]{2012ARA&A..50...29C}
---. 2012, \araa, 50, 29, \dodoi{10.1146/annurev-astro-081811-125514}

\bibitem[{{Crutcher} {et~al.}(2010){Crutcher}, {Wandelt}, {Heiles},
  {Falgarone}, \& {Troland}}]{2010ApJ...725..466C}
{Crutcher}, R.~M., {Wandelt}, B., {Heiles}, C., {Falgarone}, E., \& {Troland},
  T.~H. 2010, \apj, 725, 466, \dodoi{10.1088/0004-637X/725/1/466}

\bibitem[{{Davis} \& {Greenstein}(1951)}]{1951ApJ...114..206D}
{Davis}, Leverett, J., \& {Greenstein}, J.~L. 1951, \apj, 114, 206,
  \dodoi{10.1086/145464}

\bibitem[{{Dieleman} {et~al.}(2015){Dieleman}, {Willett}, \&
  {Dambre}}]{2015MNRAS.450.1441D}
{Dieleman}, S., {Willett}, K.~W., \& {Dambre}, J. 2015, \mnras, 450, 1441,
  \dodoi{10.1093/mnras/stv632}

\bibitem[{{Dubinski} {et~al.}(1995){Dubinski}, {Narayan}, \&
  {Phillips}}]{1995ApJ...448..226D}
{Dubinski}, J., {Narayan}, R., \& {Phillips}, T.~G. 1995, \apj, 448, 226,
  \dodoi{10.1086/175954}

\bibitem[{{Dullemond} {et~al.}(2012){Dullemond}, {Juhasz}, {Pohl}, {Sereshti},
  {Shetty}, {Peters}, {Commercon}, \& {Flock}}]{2012ascl.soft02015D}
{Dullemond}, C.~P., {Juhasz}, A., {Pohl}, A., {et~al.} 2012, {RADMC-3D: A
  multi-purpose radiative transfer tool}, Astrophysics Source Code Library,
  record ascl:1202.015.
\newblock \doeprint{1202.015}

\bibitem[{{Elmegreen} \& {Scalo}(2004)}]{2004ARA&A..42..211E}
{Elmegreen}, B.~G., \& {Scalo}, J. 2004, \araa, 42, 211,
  \dodoi{10.1146/annurev.astro.41.011802.094859}

\bibitem[{{Federrath}(2015)}]{2015MNRAS.450.4035F}
{Federrath}, C. 2015, \mnras, 450, 4035, \dodoi{10.1093/mnras/stv941}

\bibitem[{{Fosalba} {et~al.}(2002){Fosalba}, {Lazarian}, {Prunet}, \&
  {Tauber}}]{2002ApJ...564..762F}
{Fosalba}, P., {Lazarian}, A., {Prunet}, S., \& {Tauber}, J.~A. 2002, \apj,
  564, 762, \dodoi{10.1086/324297}

\bibitem[{{Goldreich} \& {Sridhar}(1995)}]{1995ApJ...438..763G}
{Goldreich}, P., \& {Sridhar}, S. 1995, \apj, 438, 763, \dodoi{10.1086/175121}

\bibitem[{{Goldsmith} {et~al.}(2008){Goldsmith}, {Heyer}, {Narayanan}, {Snell},
  {Li}, \& {Brunt}}]{2008ApJ...680..428G}
{Goldsmith}, P.~F., {Heyer}, M., {Narayanan}, G., {et~al.} 2008, \apj, 680,
  428, \dodoi{10.1086/587166}

\bibitem[{{Gonz{\'a}lez-Casanova} \& {Lazarian}(2017)}]{2017ApJ...835...41G}
{Gonz{\'a}lez-Casanova}, D.~F., \& {Lazarian}, A. 2017, \apj, 835, 41,
  \dodoi{10.3847/1538-4357/835/1/41}

\bibitem[{{Grenier} {et~al.}(2005){Grenier}, {Casandjian}, \&
  {Terrier}}]{2005Sci...307.1292G}
{Grenier}, I.~A., {Casandjian}, J.-M., \& {Terrier}, R. 2005, Science, 307,
  1292, \dodoi{10.1126/science.1106924}

\bibitem[{{Han}(2017)}]{2017ARA&A..55..111H}
{Han}, J.~L. 2017, \araa, 55, 111, \dodoi{10.1146/annurev-astro-091916-055221}

\bibitem[{He {et~al.}(2016)He, Zhang, Ren, \& Sun}]{he2016deep}
He, K., Zhang, X., Ren, S., \& Sun, J. 2016, in Proceedings of the IEEE
  conference on computer vision and pattern recognition, 770--778

\bibitem[{{Heyer} {et~al.}(2016){Heyer}, {Goldsmith}, {Y{\i}ld{\i}z}, {Snell},
  {Falgarone}, \& {Pineda}}]{2016MNRAS.461.3918H}
{Heyer}, M., {Goldsmith}, P.~F., {Y{\i}ld{\i}z}, U.~A., {et~al.} 2016, \mnras,
  461, 3918, \dodoi{10.1093/mnras/stw1567}

\bibitem[{{Heyer} {et~al.}(2020){Heyer}, {Soler}, \&
  {Burkhart}}]{2020MNRAS.496.4546H}
{Heyer}, M., {Soler}, J.~D., \& {Burkhart}, B. 2020, \mnras, 496, 4546,
  \dodoi{10.1093/mnras/staa1760}

\bibitem[{{Hoang} \& {Lazarian}(2008)}]{2008MNRAS.388..117H}
{Hoang}, T., \& {Lazarian}, A. 2008, \mnras, 388, 117,
  \dodoi{10.1111/j.1365-2966.2008.13249.x}

\bibitem[{{Hollenbach} \& {Tielens}(1999)}]{1999RvMP...71..173H}
{Hollenbach}, D.~J., \& {Tielens}, A.~G.~G.~M. 1999, Reviews of Modern Physics,
  71, 173, \dodoi{10.1103/RevModPhys.71.173}

\bibitem[{{Hu} {et~al.}(2019){Hu}, {Yuen}, {Lazarian}, {Ho}, {Benjamin},
  {Hill}, {Lockman}, {Goldsmith}, \& {Lazarian}}]{2019NatAs...3..776H}
{Hu}, Y., {Yuen}, K.~H., {Lazarian}, V., {et~al.} 2019, Nature Astronomy, 3,
  776, \dodoi{10.1038/s41550-019-0769-0}

\bibitem[{{Hutschenreuter} {et~al.}(2022){Hutschenreuter}, {Anderson}, {Betti},
  {Bower}, {Brown}, {Br{\"u}ggen}, {Carretti}, {Clarke}, {Clegg}, {Costa},
  {Croft}, {Van Eck}, {Gaensler}, {de Gasperin}, {Haverkorn}, {Heald}, {Hull},
  {Inoue}, {Johnston-Hollitt}, {Kaczmarek}, {Law}, {Ma}, {MacMahon}, {Mao},
  {Riseley}, {Roy}, {Shanahan}, {Shimwell}, {Stil}, {Sobey}, {O'Sullivan},
  {Tasse}, {Vacca}, {Vernstrom}, {Williams}, {Wright}, \&
  {En{\ss}lin}}]{2022A&A...657A..43H}
{Hutschenreuter}, S., {Anderson}, C.~S., {Betti}, S., {et~al.} 2022, \aap, 657,
  A43, \dodoi{10.1051/0004-6361/202140486}

\bibitem[{{Inoue} \& {Inutsuka}(2016)}]{2016ApJ...833...10I}
{Inoue}, T., \& {Inutsuka}, S.-i. 2016, \apj, 833, 10,
  \dodoi{10.3847/0004-637X/833/1/10}

\bibitem[{{Jansson} \& {Farrar}(2012)}]{2012ApJ...761L..11J}
{Jansson}, R., \& {Farrar}, G.~R. 2012, \apjl, 761, L11,
  \dodoi{10.1088/2041-8205/761/1/L11}

\bibitem[{{Joncas} {et~al.}(1992){Joncas}, {Boulanger}, \&
  {Dewdney}}]{1992ApJ...397..165J}
{Joncas}, G., {Boulanger}, F., \& {Dewdney}, P.~E. 1992, \apj, 397, 165,
  \dodoi{10.1086/171776}

\bibitem[{{Lazarian} \& {Vishniac}(1999)}]{1999ApJ...517..700L}
{Lazarian}, A., \& {Vishniac}, E.~T. 1999, \apj, 517, 700,
  \dodoi{10.1086/307233}

\bibitem[{{Lazarian} \& {Yuen}(2018)}]{2018ApJ...853...96L}
{Lazarian}, A., \& {Yuen}, K.~H. 2018, \apj, 853, 96,
  \dodoi{10.3847/1538-4357/aaa241}

\bibitem[{{Li} {et~al.}(2009){Li}, {Dowell}, {Goodman}, {Hildebrand}, \&
  {Novak}}]{2009ApJ...704..891L}
{Li}, H.-b., {Dowell}, C.~D., {Goodman}, A., {Hildebrand}, R., \& {Novak}, G.
  2009, \apj, 704, 891, \dodoi{10.1088/0004-637X/704/2/891}

\bibitem[{{Li} {et~al.}(2012){Li}, {Martin}, {Klein}, \&
  {McKee}}]{2012ApJ...745..139L}
{Li}, P.~S., {Martin}, D.~F., {Klein}, R.~I., \& {McKee}, C.~F. 2012, \apj,
  745, 139, \dodoi{10.1088/0004-637X/745/2/139}

\bibitem[{{Liu} {et~al.}(2022){Liu}, {Hu}, \& {Lazarian}}]{2022MNRAS.510.4952L}
{Liu}, M., {Hu}, Y., \& {Lazarian}, A. 2022, \mnras, 510, 4952,
  \dodoi{10.1093/mnras/stab3783}

\bibitem[{{Luk} {et~al.}(2022){Luk}, {Li}, \& {Li}}]{2022ApJ...928..132L}
{Luk}, S.-S., {Li}, H.-b., \& {Li}, D. 2022, \apj, 928, 132,
  \dodoi{10.3847/1538-4357/ac574c}

\bibitem[{{Mac Low}(1999)}]{1999ApJ...524..169M}
{Mac Low}, M.-M. 1999, \apj, 524, 169, \dodoi{10.1086/307784}

\bibitem[{{McKee} \& {Ostriker}(2007)}]{2007ARA&A..45..565M}
{McKee}, C.~F., \& {Ostriker}, E.~C. 2007, \araa, 45, 565,
  \dodoi{10.1146/annurev.astro.45.051806.110602}

\bibitem[{{Narayanan} {et~al.}(2008){Narayanan}, {Heyer}, {Brunt}, {Goldsmith},
  {Snell}, \& {Li}}]{2008ApJS..177..341N}
{Narayanan}, G., {Heyer}, M.~H., {Brunt}, C., {et~al.} 2008, \apjs, 177, 341,
  \dodoi{10.1086/587786}

\bibitem[{{Otto} {et~al.}(2017){Otto}, {Ji}, \& {Li}}]{2017ApJ...836...95O}
{Otto}, F., {Ji}, W., \& {Li}, H.-b. 2017, \apj, 836, 95,
  \dodoi{10.3847/1538-4357/836/1/95}

\bibitem[{{Padoan} \& {Nordlund}(1999)}]{1999ApJ...526..279P}
{Padoan}, P., \& {Nordlund}, {\r{A}}. 1999, \apj, 526, 279,
  \dodoi{10.1086/307956}

\bibitem[{{Pearson} {et~al.}(2018){Pearson}, {Palafox}, \&
  {Griffith}}]{2018MNRAS.474..478P}
{Pearson}, K.~A., {Palafox}, L., \& {Griffith}, C.~A. 2018, \mnras, 474, 478,
  \dodoi{10.1093/mnras/stx2761}

\bibitem[{{Peek} \& {Burkhart}(2019)}]{2019ApJ...882L..12P}
{Peek}, J.~E.~G., \& {Burkhart}, B. 2019, \apjl, 882, L12,
  \dodoi{10.3847/2041-8213/ab3a9e}

\bibitem[{{Planck Collaboration} {et~al.}(2011){Planck Collaboration},
  {Abergel}, {Ade}, {Aghanim}, {Arnaud}, {Ashdown}, {Aumont}, {Baccigalupi},
  {Balbi}, {Banday}, {Barreiro}, {Bartlett}, {Battaner}, {Benabed},
  {Beno{\^\i}t}, {Bernard}, {Bersanelli}, {Bhatia}, {Blagrave}, {Bock},
  {Bonaldi}, {Bond}, {Borrill}, {Bouchet}, {Boulanger}, {Bucher}, {Burigana},
  {Cabella}, {Cantalupo}, {Cardoso}, {Catalano}, {Cay{\'o}n}, {Challinor},
  {Chamballu}, {Chiang}, {Chiang}, {Christensen}, {Clements}, {Colombi},
  {Couchot}, {Coulais}, {Crill}, {Cuttaia}, {Danese}, {Davies}, {Davis}, {de
  Bernardis}, {de Gasperis}, {de Rosa}, {de Zotti}, {Delabrouille}, {Delouis},
  {D{\'e}sert}, {Dickinson}, {Donzelli}, {Dor{\'e}}, {D{\"o}rl}, {Douspis},
  {Dupac}, {Efstathiou}, {En{\ss}lin}, {Eriksen}, {Finelli}, {Forni},
  {Frailis}, {Franceschi}, {Galeotta}, {Ganga}, {Giard}, {Giardino},
  {Giraud-H{\'e}raud}, {Gonz{\'a}lez-Nuevo}, {G{\'o}rski}, {Gratton},
  {Gregorio}, {Gruppuso}, {Hansen}, {Harrison}, {Helou}, {Henrot-Versill{\'e}},
  {Herranz}, {Hildebrandt}, {Hivon}, {Hobson}, {Holmes}, {Hovest}, {Hoyland},
  {Huffenberger}, {Jaffe}, {Joncas}, {Jones}, {Jones}, {Juvela},
  {Keih{\"a}nen}, {Keskitalo}, {Kisner}, {Kneissl}, {Knox}, {Kurki-Suonio},
  {Lagache}, {Lamarre}, {Lasenby}, {Laureijs}, {Lawrence}, {Leach}, {Leonardi},
  {Leroy}, {Linden-V{\o}rnle}, {Lockman}, {L{\'o}pez-Caniego}, {Lubin},
  {Mac{\'\i}as-P{\'e}rez}, {MacTavish}, {Maffei}, {Maino}, {Mandolesi}, {Mann},
  {Maris}, {Marshall}, {Martin}, {Mart{\'\i}nez-Gonz{\'a}lez}, {Masi},
  {Matarrese}, {Matthai}, {Mazzotta}, {McGehee}, {Meinhold}, {Melchiorri},
  {Mendes}, {Mennella}, {Miville-Desch{\^e}nes}, {Moneti}, {Montier},
  {Morgante}, {Mortlock}, {Munshi}, {Murphy}, {Naselsky}, {Nati}, {Natoli},
  {Netterfield}, {N{\o}rgaard-Nielsen}, {Noviello}, {Novikov}, {Novikov},
  {O'Dwyer}, {Osborne}, {Pajot}, {Paladini}, {Pasian}, {Patanchon},
  {Perdereau}, {Perotto}, {Perrotta}, {Piacentini}, {Piat}, {Pinheiro
  Gon{{c}}alves}, {Plaszczynski}, {Pointecouteau}, {Polenta}, {Ponthieu},
  {Poutanen}, {Pr{\'e}zeau}, {Prunet}, {Puget}, {Rachen}, {Reach}, {Reinecke},
  {Renault}, {Ricciardi}, {Riller}, {Ristorcelli}, {Rocha}, {Rosset},
  {Rowan-Robinson}, {Rubi{\~n}o-Mart{\'\i}n}, {Rusholme}, {Sandri}, {Santos},
  {Savini}, {Scott}, {Seiffert}, {Shellard}, {Smoot}, {Starck}, {Stivoli},
  {Stolyarov}, {Stompor}, {Sudiwala}, {Sygnet}, {Tauber}, {Terenzi},
  {Toffolatti}, {Tomasi}, {Torre}, {Tristram}, {Tuovinen}, {Umana},
  {Valenziano}, {Vielva}, {Villa}, {Vittorio}, {Wade}, {Wandelt}, {Wilkinson},
  {Yvon}, {Zacchei}, \& {Zonca}}]{2011A&A...536A..24P}
{Planck Collaboration}, {Abergel}, A., {Ade}, P.~A.~R., {et~al.} 2011, \aap,
  536, A24, \dodoi{10.1051/0004-6361/201116485}

\bibitem[{{Planck Collaboration} {et~al.}(2016){Planck Collaboration}, {Ade},
  {Aghanim}, {Alves}, {Arnaud}, {Arzoumanian}, {Ashdown}, {Aumont},
  {Baccigalupi}, {Banday}, {Barreiro}, {Bartolo}, {Battaner}, {Benabed},
  {Beno{\^\i}t}, {Benoit-L{\'e}vy}, {Bernard}, {Bersanelli}, {Bielewicz},
  {Bock}, {Bonavera}, {Bond}, {Borrill}, {Bouchet}, {Boulanger}, {Bracco},
  {Burigana}, {Calabrese}, {Cardoso}, {Catalano}, {Chiang}, {Christensen},
  {Colombo}, {Combet}, {Couchot}, {Crill}, {Curto}, {Cuttaia}, {Danese},
  {Davies}, {Davis}, {de Bernardis}, {de Rosa}, {de Zotti}, {Delabrouille},
  {Dickinson}, {Diego}, {Dole}, {Donzelli}, {Dor{\'e}}, {Douspis}, {Ducout},
  {Dupac}, {Efstathiou}, {Elsner}, {En{\ss}lin}, {Eriksen},
  {Falceta-Gon{{c}}alves}, {Falgarone}, {Ferri{\`e}re}, {Finelli}, {Forni},
  {Frailis}, {Fraisse}, {Franceschi}, {Frejsel}, {Galeotta}, {Galli}, {Ganga},
  {Ghosh}, {Giard}, {Gjerl{\o}w}, {Gonz{\'a}lez-Nuevo}, {G{\'o}rski},
  {Gregorio}, {Gruppuso}, {Gudmundsson}, {Guillet}, {Harrison}, {Helou},
  {Hennebelle}, {Henrot-Versill{\'e}}, {Hern{\'a}ndez-Monteagudo}, {Herranz},
  {Hildebrandt}, {Hivon}, {Holmes}, {Hornstrup}, {Huffenberger}, {Hurier},
  {Jaffe}, {Jaffe}, {Jones}, {Juvela}, {Keih{\"a}nen}, {Keskitalo}, {Kisner},
  {Knoche}, {Kunz}, {Kurki-Suonio}, {Lagache}, {Lamarre}, {Lasenby},
  {Lattanzi}, {Lawrence}, {Leonardi}, {Levrier}, {Liguori}, {Lilje},
  {Linden-V{\o}rnle}, {L{\'o}pez-Caniego}, {Lubin}, {Mac{\'\i}as-P{\'e}rez},
  {Maino}, {Mandolesi}, {Mangilli}, {Maris}, {Martin},
  {Mart{\'\i}nez-Gonz{\'a}lez}, {Masi}, {Matarrese}, {Melchiorri}, {Mendes},
  {Mennella}, {Migliaccio}, {Miville-Desch{\^e}nes}, {Moneti}, {Montier},
  {Morgante}, {Mortlock}, {Munshi}, {Murphy}, {Naselsky}, {Nati},
  {Netterfield}, {Noviello}, {Novikov}, {Novikov}, {Oppermann}, {Oxborrow},
  {Pagano}, {Pajot}, {Paladini}, {Paoletti}, {Pasian}, {Perotto}, {Pettorino},
  {Piacentini}, {Piat}, {Pierpaoli}, {Pietrobon}, {Plaszczynski},
  {Pointecouteau}, {Polenta}, {Ponthieu}, {Pratt}, {Prunet}, {Puget}, {Rachen},
  {Reinecke}, {Remazeilles}, {Renault}, {Renzi}, {Ristorcelli}, {Rocha},
  {Rossetti}, {Roudier}, {Rubi{\~n}o-Mart{\'\i}n}, {Rusholme}, {Sandri},
  {Santos}, {Savelainen}, {Savini}, {Scott}, {Soler}, {Stolyarov}, {Sudiwala},
  {Sutton}, {Suur-Uski}, {Sygnet}, {Tauber}, {Terenzi}, {Toffolatti}, {Tomasi},
  {Tristram}, {Tucci}, {Umana}, {Valenziano}, {Valiviita}, {Van Tent},
  {Vielva}, {Villa}, {Wade}, {Wandelt}, {Wehus}, {Ysard}, {Yvon}, \&
  {Zonca}}]{2016A&A...586A.138P}
{Planck Collaboration}, {Ade}, P.~A.~R., {Aghanim}, N., {et~al.} 2016, \aap,
  586, A138, \dodoi{10.1051/0004-6361/201525896}

\bibitem[{{Planck Collaboration} {et~al.}(2020){Planck Collaboration},
  {Aghanim}, {Akrami}, {Alves}, {Ashdown}, {Aumont}, {Baccigalupi},
  {Ballardini}, {Banday}, {Barreiro}, {Bartolo}, {Basak}, {Benabed}, {Bernard},
  {Bersanelli}, {Bielewicz}, {Bock}, {Bond}, {Borrill}, {Bouchet}, {Boulanger},
  {Bracco}, {Bucher}, {Burigana}, {Calabrese}, {Cardoso}, {Carron}, {Chary},
  {Chiang}, {Colombo}, {Combet}, {Crill}, {Cuttaia}, {de Bernardis}, {de
  Zotti}, {Delabrouille}, {Delouis}, {Di Valentino}, {Dickinson}, {Diego},
  {Dor{\'e}}, {Douspis}, {Ducout}, {Dupac}, {Efstathiou}, {Elsner},
  {En{\ss}lin}, {Eriksen}, {Falgarone}, {Fantaye}, {Fernandez-Cobos},
  {Ferri{\`e}re}, {Finelli}, {Forastieri}, {Frailis}, {Fraisse}, {Franceschi},
  {Frolov}, {Galeotta}, {Galli}, {Ganga}, {G{\'e}nova-Santos}, {Gerbino},
  {Ghosh}, {Gonz{\'a}lez-Nuevo}, {G{\'o}rski}, {Gratton}, {Green}, {Gruppuso},
  {Gudmundsson}, {Guillet}, {Handley}, {Hansen}, {Helou}, {Herranz}, {Hivon},
  {Huang}, {Jaffe}, {Jones}, {Keih{\"a}nen}, {Keskitalo}, {Kiiveri}, {Kim},
  {Krachmalnicoff}, {Kunz}, {Kurki-Suonio}, {Lagache}, {Lamarre}, {Lasenby},
  {Lattanzi}, {Lawrence}, {Le Jeune}, {Levrier}, {Liguori}, {Lilje},
  {Lindholm}, {L{\'o}pez-Caniego}, {Lubin}, {Ma}, {Mac{\'\i}as-P{\'e}rez},
  {Maggio}, {Maino}, {Mandolesi}, {Mangilli}, {Marcos-Caballero}, {Maris},
  {Martin}, {Mart{\'\i}nez-Gonz{\'a}lez}, {Matarrese}, {Mauri}, {McEwen},
  {Melchiorri}, {Mennella}, {Migliaccio}, {Miville-Desch{\^e}nes}, {Molinari},
  {Moneti}, {Montier}, {Morgante}, {Moss}, {Natoli}, {Pagano}, {Paoletti},
  {Patanchon}, {Perrotta}, {Pettorino}, {Piacentini}, {Polastri}, {Polenta},
  {Puget}, {Rachen}, {Reinecke}, {Remazeilles}, {Renzi}, {Ristorcelli},
  {Rocha}, {Rosset}, {Roudier}, {Rubi{\~n}o-Mart{\'\i}n}, {Ruiz-Granados},
  {Salvati}, {Sandri}, {Savelainen}, {Scott}, {Sirignano}, {Sunyaev},
  {Suur-Uski}, {Tauber}, {Tavagnacco}, {Tenti}, {Toffolatti}, {Tomasi},
  {Trombetti}, {Valiviita}, {Vansyngel}, {Van Tent}, {Vielva}, {Villa},
  {Vittorio}, {Wandelt}, {Wehus}, {Zacchei}, \& {Zonca}}]{2020A&A...641A..12P}
{Planck Collaboration}, {Aghanim}, N., {Akrami}, Y., {et~al.} 2020, \aap, 641,
  A12, \dodoi{10.1051/0004-6361/201833885}

\bibitem[{{Rao} {et~al.}(1998){Rao}, {Crutcher}, {Plambeck}, \&
  {Wright}}]{1998ApJ...502L..75R}
{Rao}, R., {Crutcher}, R.~M., {Plambeck}, R.~L., \& {Wright}, M.~C.~H. 1998,
  \apjl, 502, L75, \dodoi{10.1086/311485}

\bibitem[{{Ridge} {et~al.}(2006){Ridge}, {Di Francesco}, {Kirk}, {Li},
  {Goodman}, {Alves}, {Arce}, {Borkin}, {Caselli}, {Foster}, {Heyer},
  {Johnstone}, {Kosslyn}, {Lombardi}, {Pineda}, {Schnee}, \&
  {Tafalla}}]{2006AJ....131.2921R}
{Ridge}, N.~A., {Di Francesco}, J., {Kirk}, H., {et~al.} 2006, \aj, 131, 2921,
  \dodoi{10.1086/503704}

\bibitem[{Ronneberger {et~al.}(2015)Ronneberger, Fischer, \&
  Brox}]{ronneberger2015u}
Ronneberger, O., Fischer, P., \& Brox, T. 2015, in International Conference on
  Medical image computing and computer-assisted intervention, Springer,
  234--241

\bibitem[{{Soler}(2019)}]{2019A&A...629A..96S}
{Soler}, J.~D. 2019, \aap, 629, A96, \dodoi{10.1051/0004-6361/201935779}

\bibitem[{{Soler} \& {Hennebelle}(2017)}]{2017A&A...607A...2S}
{Soler}, J.~D., \& {Hennebelle}, P. 2017, \aap, 607, A2,
  \dodoi{10.1051/0004-6361/201731049}

\bibitem[{{Soler} {et~al.}(2013){Soler}, {Hennebelle}, {Martin},
  {Miville-Desch{\^e}nes}, {Netterfield}, \& {Fissel}}]{2013ApJ...774..128S}
{Soler}, J.~D., {Hennebelle}, P., {Martin}, P.~G., {et~al.} 2013, \apj, 774,
  128, \dodoi{10.1088/0004-637X/774/2/128}

\bibitem[{{Soler} {et~al.}(2017){Soler}, {Ade}, {Angil{\`e}}, {Ashton},
  {Benton}, {Devlin}, {Dober}, {Fissel}, {Fukui}, {Galitzki}, {Gandilo},
  {Hennebelle}, {Klein}, {Li}, {Korotkov}, {Martin}, {Matthews}, {Moncelsi},
  {Netterfield}, {Novak}, {Pascale}, {Poidevin}, {Santos}, {Savini}, {Scott},
  {Shariff}, {Thomas}, {Tucker}, {Tucker}, \&
  {Ward-Thompson}}]{2017A&A...603A..64S}
{Soler}, J.~D., {Ade}, P.~A.~R., {Angil{\`e}}, F.~E., {et~al.} 2017, \aap, 603,
  A64, \dodoi{10.1051/0004-6361/201730608}

\bibitem[{{Soler} {et~al.}(2019){Soler}, {Beuther}, {Rugel}, {Wang}, {Clark},
  {Glover}, {Goldsmith}, {Heyer}, {Anderson}, {Goodman}, {Henning},
  {Kainulainen}, {Klessen}, {Longmore}, {McClure-Griffiths}, {Menten},
  {Mottram}, {Ott}, {Ragan}, {Smith}, {Urquhart}, {Bigiel}, {Hennebelle},
  {Roy}, \& {Schilke}}]{2019A&A...622A.166S}
{Soler}, J.~D., {Beuther}, H., {Rugel}, M., {et~al.} 2019, \aap, 622, A166,
  \dodoi{10.1051/0004-6361/201834300}

\bibitem[{{Troland} \& {Heiles}(1986)}]{1986ApJ...301..339T}
{Troland}, T.~H., \& {Heiles}, C. 1986, \apj, 301, 339, \dodoi{10.1086/163904}

\bibitem[{{Umemoto} {et~al.}(2017){Umemoto}, {Minamidani}, {Kuno}, {Fujita},
  {Matsuo}, {Nishimura}, {Torii}, {Tosaki}, {Kohno}, {Kuriki}, {Tsuda},
  {Hirota}, {Ohashi}, {Yamagishi}, {Handa}, {Nakanishi}, {Omodaka}, {Koide},
  {Matsumoto}, {Onishi}, {Tokuda}, {Seta}, {Kobayashi}, {Tachihara}, {Sano},
  {Hattori}, {Onodera}, {Oasa}, {Kamegai}, {Tsuboi}, {Sofue}, {Higuchi},
  {Chibueze}, {Mizuno}, {Honma}, {Muller}, {Inoue}, {Morokuma-Matsui},
  {Shinnaga}, {Ozawa}, {Takahashi}, {Yoshiike}, {Costes}, \&
  {Kuwahara}}]{2017PASJ...69...78U}
{Umemoto}, T., {Minamidani}, T., {Kuno}, N., {et~al.} 2017, \pasj, 69, 78,
  \dodoi{10.1093/pasj/psx061}

\bibitem[{{Van Oort} {et~al.}(2019){Van Oort}, {Xu}, {Offner}, \&
  {Gutermuth}}]{2019ApJ...880...83V}
{Van Oort}, C.~M., {Xu}, D., {Offner}, S. S.~R., \& {Gutermuth}, R.~A. 2019,
  \apj, 880, 83, \dodoi{10.3847/1538-4357/ab275e}

\bibitem[{{Xu} {et~al.}(2016){Xu}, {Li}, {Yue}, \&
  {Goldsmith}}]{2016ApJ...819...22X}
{Xu}, D., {Li}, D., {Yue}, N., \& {Goldsmith}, P.~F. 2016, \apj, 819, 22,
  \dodoi{10.3847/0004-637X/819/1/22}

\bibitem[{{Xu} {et~al.}(2022){Xu}, {Offner}, {Gutermuth}, {Kong}, \&
  {Arce}}]{2022ApJ...926...19X}
{Xu}, D., {Offner}, S. S.~R., {Gutermuth}, R., {Kong}, S., \& {Arce}, H.~G.
  2022, \apj, 926, 19, \dodoi{10.3847/1538-4357/ac39a0}

\bibitem[{{Xu} {et~al.}(2020{\natexlab{a}}){Xu}, {Offner}, {Gutermuth}, \&
  {Oort}}]{2020ApJ...890...64X}
{Xu}, D., {Offner}, S. S.~R., {Gutermuth}, R., \& {Oort}, C.~V.
  2020{\natexlab{a}}, \apj, 890, 64, \dodoi{10.3847/1538-4357/ab6607}

\bibitem[{{Xu} {et~al.}(2020{\natexlab{b}}){Xu}, {Offner}, {Gutermuth}, \&
  {Oort}}]{2020ApJ...905..172X}
---. 2020{\natexlab{b}}, \apj, 905, 172, \dodoi{10.3847/1538-4357/abc7bf}

\bibitem[{{Yuen} \& {Lazarian}(2017)}]{2017ApJ...837L..24Y}
{Yuen}, K.~H., \& {Lazarian}, A. 2017, \apjl, 837, L24,
  \dodoi{10.3847/2041-8213/aa6255}

\bibitem[{{Zhang} {et~al.}(2019){Zhang}, {Guo}, {Wang}, \&
  {Li}}]{2019ApJ...871...98Z}
{Zhang}, Y., {Guo}, Z., {Wang}, H.~H., \& {Li}, H.~b. 2019, \apj, 871, 98,
  \dodoi{10.3847/1538-4357/aaf57c}

\end{thebibliography}

\end{CJK*}

\end{document}